\documentclass[12pt,preprint]{aastex} 
\usepackage{emulateapj5} 

\slugcomment{Accepted by ApJ}

\shortauthors{McMahon, White, Becker, \& Helfand}
\shorttitle{Optical IDs for 70,000 Radio Sources}

\begin{document}

\title{Optical Counterparts for 70,000 Radio Sources: APM
Identifications for the {\it FIRST} Radio Survey}

\author{Richard G. McMahon}
\affil{University of Cambridge, Institute of Astronomy, Madingley Road,
Cambridge CB3 0HA, England}
\email{rgm@ast.cam.ac.uk}

\author{Richard L. White}
\affil{Space Telescope Science Institute,
3700 San Martin Dr., Baltimore, MD 21218}
\email{rlw@stsci.edu}

\author{David J. Helfand}
\affil{Astronomy Dept., Columbia University, New York, NY 10027}
\email{djh@astro.columbia.edu}

\author{Robert H. Becker}
\affil{Physics Dept., University of California, Davis, CA  95616\\
and IGPP/Lawrence Livermore National Laboratory}
\email{bob@igpp.ucllnl.org}

\begin{abstract}
  
  We describe a program to identify optical counterparts to radio
  sources from the VLA {\it FIRST} survey using the Cambridge APM
  scans of the POSS-I plates. We use radio observations covering 4150
  deg$^2$ of the north Galactic cap to a 20~cm flux density threshold
  of 1.0~mJy; the 382,892 sources detected all have positional
  uncertainties of $<1^{\prime\prime}$ (radius of 90\% confidence).
  Our description of the APM catalog, derived from the 148 POSS-I $O$
  and $E$ plates covering this region, includes an assessment of its
  astrometric and photometric accuracy, a photometric recalibration
  using the Minnesota APS catalog, a discussion of the classification
  algorithm, and quantitative tests of the catalog's reliability and
  completeness. We go on to show how the use of {\it FIRST} sources as
  astrometric standards allows us to improve the absolute astrometry
  of the POSS plates by nearly an order of magnitude to $\sim
  0.15^{\prime \prime}$ rms. Matching the radio and optical catalogs
  yields counterparts for over 70,000 radio sources; we include
  detailed discussions of the reliability and completeness of these
  identifications as a function of optical and radio morphology,
  optical magnitude and color, and radio flux density. An analysis of
  the problem of radio sources with complex morphologies ({\it e.g.},
  double-lobed radio galaxies) is included. We conclude with a brief
  discussion of the source classes represented among the radio sources
  with identified counterparts.

\end{abstract}


\section{Introduction}
\label{sectionintro}

The optical identification of the first discrete extraterrestrial radio
source occurred as a result of a telephone call from J.~S. Hey to the
Royal Greenwich Observatory on the afternoon of February 28, 1942.
Recognizing that the source of extensive jamming of British radar over
the previous two days appeared to follow the Sun, Hey was delighted to
learn that an unusually large sunspot had just transited the solar
disk; despite its skeptical reception by his superiors, Hey's
identification proved correct (Hey 1973).

In the ensuing decade, progress in the detection of new extrasolar
radio emitters far outstripped the ability of astronomers to associate
them with optical counterparts. The first breakthrough came in 1949
when Bolton, Stanley and Slee (1949) identified the Crab Nebula, M87,
and NGC 5128 (Cen A) with three of the brightest radio sources in the
sky, although they concluded that the bizarre morphology of the latter
generally favored a Galactic interpretation for radio emitters since
``the probability of [such] an unusual object in our own Galaxy seems
greater than a large accumulation of such objects at a great
distance.'' The following year, Ryle, Smith, and Elsmore (1950)
concurred in this conclusion despite finding 0/146 bright ($V < $4.0)
stars, 0/21 novae, 0/38 planetary nebulae, 0/29 diffuse Galactic
nebulae, and 4/5 of the brightest galaxies coincident with entries in
their fifty-source radio catalog. It was not until the classic papers
of Baade \& Minkowski (1954a,b), which among other things pronounced
Cygnus A ``an extragalactic affair'', that the era of extragalactic
radio source identification can be said to have begun.

The largest radio catalogs in existence prior to 1995 contained, in
total, approximately 100,000 distinct entries. In striking contrast to
the earliest speculations, fewer than 20 of these relatively bright
radio sources are identified as stars. Indeed, fewer than 1000 stellar
radio detections have been made despite decades of sensitive, targeted
searches (Hjellming 1988 and references therein; Wendker 1995), and
$<5\%$ of all cataloged radio sources are Galactic objects. A search of
the NED database, however, suggests that the fraction of identified
{\it extragalactic} radio emitters today is little better than it was
in 1950, when 7 out of 67 known radio sources had identified
counterparts (Baade and Minkowski 1954b). The problem now is the same
as it was 50 years ago:  the angular resolution of large-area radio
surveys is generally too poor (several arcminutes) to allow for the
unambiguous association of cataloged objects with individual optical
counterparts which, at the limit of the POSS-I plates, number $>$4000
deg$^{-2}$ at high Galactic latitudes. Interferometric surveys at
centimeter wavelengths can achieve the desired positional accuracy of
$\sim 1\arcsec$ but have, until recently, covered only $\sim50$~deg$^{2}$ of
sky, resulting in fewer than 500 optical identifications
for radio-selected objects at faint flux levels (see Table~\ref{table-ids}).

While considerable information on the class and emission mechanism of a
radio source can be derived from observations of its radio morphology, its
spectrum, and its polarization characteristics, optical observations are
still required to establish the source's distance and to classify it
unambiguously. Eight years ago, we began to construct Faint Images of the
Radio Sky at Twenty-cm ({\it FIRST}) with the primary goal of obtaining a very
large sample of radio sources with positions sufficiently accurate that
the majority of objects detected could be easily identified on the
basis of positional coincidence alone. {\it FIRST} has
been designed to cover the same 10,000 deg$^{2}$ region as the Sloan
Digital Sky Survey (SDSS), which will obtain deep optical images of the northern
sky in five colors, and will take spectra of a million objects over the
next decade (Gunn and Knapp 1993). However, an archive of the optical sky
in the form of the National Geographic-Palomar Observatory Sky Survey
(POSS) plates already exists. The identification of even
$\sim$20\% of all {\it FIRST} sources with counterparts at or above the POSS
plate limit will immediately provide samples of various radio source
populations from one to three orders of magnitude larger than those in
existence, and will advance significantly our knowledge of the radio
universe.

In this paper, we present the results of an optical identification
program for {\it FIRST} radio sources based on the Cambridge Automated
Plate Measuring Machine (APM) scans of the POSS I plates (McMahon and
Irwin 1992). In \S\ref{sectionsurvey}, we describe briefly the
parameters of the {\it FIRST} survey and the catalog derived
therefrom. In \S\ref{sectionapm}, we describe the genesis of the APM
catalog and discuss its astrometric, photometric, and source
classification uncertainties with particular attention to those which
either 1) the {\it FIRST} results can help to refine, or 2) are
particularly relevant to radio source identification; the following
section (\S\ref{sectionastrometry}) demonstrates the utility of {\it
  FIRST} as an astrometric calibrator by deriving both intraplate and
plate-to-plate corrections to the optical astrometric solution.
Section~\ref{sectionids} describes the matching algorithms employed
and discusses such issues as false match rates, reliability and
completeness of the proposed identifications, and the effects of
optical and radio source morphology. Section 6 discusses color,
magnitude, and radio flux density distributions for the more than
70,000 counterparts we have identified, while the following section
(\S\ref{sectioncatalog}) describes the format of the counterpart
catalog (the full contents of which are available on the Web).  In the
discussion (\S\ref{sectiondiscussion}), we place in context this
identification program for faint radio sources, and summarize the
statistics of the populations represented. A reprise of our results
and a precis of future work concludes our report
(\S\ref{sectionsummary}).

\bigskip \bigskip \bigskip 

\section{The FIRST Survey}
\label{sectionsurvey}

In its initial five years, the {\it FIRST} project surveyed
$\sim$4670~deg$^{2}$ of the high Galactic latitude sky to a point-source
flux density limit of 1.0 mJy at a wavelength of 20~cm.  Using the
B-configuration of the Very Large Array (VLA\footnote{The National Radio
Astronomy Observatory is operated by Associated Universities, Inc., under
cooperative agreement with the National Science Foundation.}) in
bandwidth-synthesis mode, we have produced nearly uniform-sensitivity images of
the sky with a median rms of 0.14 mJy and an angular resolution of 5\arcsec;
the positions for each of the half million sources in our current catalog
are accurate to better than 1\arcsec\ (90\% confidence). Details of the
observing strategy and the pipeline processing system that creates the
final images from the raw uv data are presented in Becker, White, and
Helfand (1995; hereafter, BWH).

For the program described here, we have utilized the 98Feb04 version of
the {\it FIRST} catalog,
which is available from our Web homepage
(\url{http://sundog.stsci.edu}).  The
catalog contains positions, peak and integrated flux densities, source
morphological parameters, and information on the field from which each entry was
derived. A detailed description of the catalog's construction can be found
in White et al. (1997; hereafter WBHG); an exhaustive discussion of the
astrometric, photometric, and morphological uncertainties is included in
that paper and in BWH, and is summarized on our Web homepage.

The catalog has a mean surface density of $\sim$90 ``sources'' per
square degree. However, it is important to note that a significant
fraction of radio sources at these flux densities are extended. In
particular, classical double-lobed radio sources in which the
components are separated by more than $\sim$3\arcsec\ are typically
represented by two catalog entries; for complex sources extended on
scales of $\sim$1\arcmin, several components can be required to fit
adequately the source surface brightness distribution (see WBHG). We
have not yet completed a definitive study of multi-component sources;
even when this is done, it will not be possible to decide unambiguously
whether or not two nearby catalog entries are parts of the same source
hosted by a single optical counterpart, or whether they are chance
alignments of unrelated objects. We briefly investigate these issues
in \S5.4; prior to that discussion, we use the term ``sources''
synonymously with ``radio catalog entries''.

For clarity, we present in this paper only the results of the APM
identification program for the $\sim$~4150 deg$^2$ in the north Galactic
cap. This region has approximate boundaries of
$+22^{\circ}<\delta<+58^{\circ}$ and $07^{\rm h}30^{\rm m}<RA<17^{\rm h}30^{\rm m}$ and
includes $\sim$~382,892 radio sources. Subsequent extensions to this
work for the two southern cap strips
($+1.6^{\circ}>\delta<-2.5^{\circ}$ and $-7^{\circ}>\delta>-11^{\circ}$
in the range $21^{\rm h}30^{\rm m}<RA<03^{\rm h}00^{\rm m}$ for a total of 610 deg$^2$) as well
as the remainder of the north cap region will be added to the Web site
as they become available.

\section{The APM Catalog}
\label{sectionapm}

\subsection{Introduction}

It has been almost 50 years since the National Geographic
Society-Palomar Observatory Sky Survey (POSS) was carried out
(1949 to 1958) using the $48^{\prime\prime}$ Oschin Schmidt at Palomar
Observatory 
(Harrington 1952; Minkowski and Abell 1963; Lund and Dixon 1973).  Since a
second-epoch Palomar Survey is underway, the original POSS is now
usually referred to as the first-epoch POSS or POSS-I to differentiate
it from POSS-II (Reid et al. 1991). The survey covers the
31,000 deg$^2$ of sky north of $\delta\sim-33^\circ$ with
6.4$\times$6.4 degree plates in two colors: blue ($O$), centered
at $\sim4100$~\AA\ covering the range between 4900~\AA\ and the atmospheric 
cutoff at $\sim$3300,
and red ($E$), a 600~\AA\ wide band centered at $\sim$6500~\AA.
Plots of the POSS-I bandpasses compared with the standard Johnson
photometric system can be seen in Evans (1989). The nearest
Johnson system bands to O and E are B and R respectively. We
provide color equations for these bands in section~\ref{sectionApmPhotometry}.

Between 1991 and 1995, the SERC Automated Plate Measuring Machine (APM)
at the Institute for Astronomy, Cambridge (Kibblewhite et al. 1984) was
used to digitize these plates at a resolution of 7.5 microns per
pixel (0.5\arcsec), the highest spatial resolution yet applied to these
images (McMahon and Irwin 1992). An object catalog has been constructed
from these data which includes all objects down to the plate limits ---
$\sim$20.0 in $E$ and $\sim$21.5 in $O$ --- and contains approximately 2000
stars and 2000 galaxies deg$^{-2}$ at high Galactic latitudes. The
catalog contains positions, magnitudes, morphological classification
parameters, major and minor axes, and isophotal areas for each source;
a merged catalog which matches objects between plates also contains a
color (or an upper limit thereto) for each entry. An automated
classification algorithm interprets the morphological parameters to
classify each object.  Here, we present the basic procedures which
establish the astrometric and photometric calibration of the APM
catalog, and discuss the limits of the image classification system. 

\subsection{APM Astrometry}

The APM machine measures the $x$ and $y$ positions of all objects
detected.  The conversion relationship between these measured APM
positions and celestial coordinates is derived by matching stars in the
Tycho-ACT catalog (Hog et al. 1997; 
Urban, Corbin \& Wycoff 1998) with stars detected on each plate using
a `standard' six plate-constant model that allows for shift, rotation,
scale, and shear. The algorithm uses iterative $2.5\sigma$ 
clipping to give a
robust fit; the typical rms on the fitted positions of the Tycho stars
are $0.4^{\prime\prime}$--$0.8^{\prime\prime}$.

Irwin (1994) has studied the two-dimensional systematic errors in an
earlier version of the APM catalog positions by investigating the
intraplate residuals between the measured positions for bright stars
in the Positions and Proper Motions Catalog (Roser and Bastian 1991)
and the astrometric fit.  He found significant, systematic residuals
ranging up to $0.5^{\prime\prime}$. In the version of the APM
catalogue used herein, the astrometric analysis uses the more recent
ACT catalog (Urban, Corbin \& Wycoff 1998).

A residual map generated from this analysis
is applied to positions in the standard APM catalog available at
\url{http://www.ast.cam.ac.uk/\~{}apmcat}.  We discuss below (\S 4)
the significant
improvement in overall astrometric accuracy that can be derived from a
comparison of radio source counterparts in the APM with {\it FIRST}
survey source positions.

\subsection{APM Photometry}
\label{sectionApmPhotometry}

The APM measures photographic density rather than flux; moreover, the
central regions of all objects more than a factor of $\sim$10 brighter
than the sky produce a nonlinear response and/or are saturated. The
algorithms used to overcome these inherent difficulties are discussed
in detail by Irwin (1985). Briefly, a local background is determined
for each of $\sim$500000 locations on each plate by producing a
histogram of the pixel values in 64 $\times$ 64-pixel regions
($32^{\prime\prime} \times 32^{\prime\prime}$) and finding the mode of
each distribution; 
two-dimensional smoothing is applied to these half million background
estimates to derive a background model for the plate. The image
detection algorithm then finds connected regions of pixels above a
threshold level (typically $2\sigma$ above the estimated background
level for the given plate position).  This background-following
technique has the advantage that faint objects lying in the halos of
bright objects can be detected. However, large objects such as bright
stars and galaxies with angular extents $>30^{\prime\prime}$ have
their raw fluxes underestimated. An additional problem for large
images is that the limited memory available to the software means that
bright objects sometimes overflow the pixel buffers and are lost. This
occurs for images with sizes greater than roughly 1--2~mm ({\it i.e.},
1--$2^{\prime}$), corresponding to stellar magnitudes brighter than
$\sim 9.0$.

Another inherent problem arises in attempting to derive magnitude estimates for
extended objects from saturated images. Saturation effects can be corrected for
in stars by assuming that stellar images have an intrinsic density profile 
independent of magnitude, and that this profile can be derived from the
unsaturated parts of stellar profiles. A high signal-to-noise intrinsic profile is constructed by taking the core from faint stars and the wings from 
brighter stars (see Bunclark \& Irwin (1983) for further details).
This profile can then be integrated and used to derive a calibration 
curve to convert saturated stellar magnitudes to a linear
system. In the APM catalog, this calibration is applied to
all images. This has the unfortunate consequence that galaxies, which
have shallower surface brightness profiles and lower central surface
brightnesses than stars of the same
total magnitude, will have their magnitudes over-corrected. This is
a fundamental problem for galaxy photometry determined from
photographic sky survey plates (see Metcalfe, Fong, and Shanks (1995) for
a discussion).

The basic APM catalog is {\it defined} to have a red-band ($E$) plate
limit of $m({\rm R_e})=20.0$. This limit was established during the early
stages ($\sim$1991) of the creation of the APM catalog via comparison
with $\sim$10 photometric sequences (Evans 1989; Humphreys
et al. 1991).  Similarly, a single slope of 1.10, was assumed in
converting between the linearized APM magnitudes (Bunclark \& Irwin
1983) and the $\alpha$ Lyrae-based Johnson magnitude system.  It was
noted at the time that there were significant deviations ($\sim$1~mag)
from a simple linear relation at magnitudes brighter than $\sim$15.
This is not surprising bearing in mind that the POSS-I glass plates
measured by the APM are copies that may have different degrees of
saturation and have had their contrast stretched to enhance faint
features. The assumption of a constant flux limit seemed reasonable,
since the plates were all taken in similar dark sky observing
conditions with exposure times that were adjusted to ensure uniform
sensitivity. A similar assumption is made in all modern photographic
cameras where it is assumed that all photographic film has the
specified speed.  The blue band ($O$) limit was defined with respect to
the red limit; for the 428 fields available in March 1999, this has
a range of $m(\rm B_o$)=20.6--21.3 ($\pm$1$\sigma$).

Eventually, a full photometric recalibration of the APM using the
Guide Star Photometric Catalog (GSPC -- Postman et al. 1998a) CCD
sequences is planned.  Preliminary comparisons with CCD photometry for
$\sim5$\% of the POSS-I plates show that the APM magnitudes for
stellar objects have a global rms uncertainty of 0.5 magnitudes over
the range 16 to 20, the range in which most {\it FIRST} counterparts
lie.  As discussed above, the uncertainties in the magnitudes of
galaxies are more complex, since galaxies have a range of surface
brightness distributions, and hence may have complex, partially
saturated surface brightness profiles on the POSS-I plates. This is
compounded by the range in calibration slopes observed. At faint
magnitudes (18--20) where the image profiles are unsaturated, the APM
magnitudes may be more reliable, but it is left to the reader to
verify this where precise magnitudes are required. For many programs,
a uniform set of magnitudes or uniform selection criteria are more
critical.

It is also worth noting that almost 50 years has elapsed between the
epochs of the POSS-I and {\it FIRST} surveys, so that optical
variability is an additional uncertainty. Hook et al. (1994) have
studied the long term variability of radio quiet quasars and found that
over a rest-frame period of $\sim$10 years, a typical $m(B)=19$ quasar
varies by 0.20 magnitudes (rms). The longer-term variability of {\it
FIRST} optical counterparts could be studied via a comparison between
the POSS-I plates and the POSS-II or UKST plates. A CCD investigation
of the long-term variability of $\sim200$ quasars from the {\it FIRST}
Bright Quasar Survey (White et al. 2000) is reported elsewhere
(Helfand et al. 2001).

It is useful to be able to convert the
magnitudes of the O and E bands to the nearest bands in the Johnson
Vega-normalised magnitude system.  Evans (1989) has found that over the color
range 0.0 $<$ B$-$R $<$ 1.5,

\begin{equation}
 R-E = (0.00\pm0.02)   (B-R)
\end{equation}

Thus one can assume;

\begin{equation}
    R=E
\end{equation}

Evans (1989) also found

\begin{equation}
O-E = (1.135\pm0.035) (B-R)
\end{equation}

Thus we have;

\begin{equation}
    B-R = 0.88(O-E)
\end{equation}

Finally it follows from (1) and (3) that:

\begin{equation}
    B = O-0.12(O-E)
\end{equation}

Assuming central wavelengths of 4100\AA\ and 6500\AA\ and the Hayes
and Latham (1975) calibration of Vega the monochromatic zero point of
the O and E bands, i.e., the flux corresponding to a magnitude of zero,
is 4550~Jy and 2980~Jy respectively.  Caution is advised when using
these conversions since the presence of emission lines in the O and E
filters will effect the conversion.

\subsection{Photometric Calibration Using APS}

In an attempt to improve the APM photometric accuracy and uniformity,
the APM magnitudes in regions covered by the {\it FIRST} survey
were recalibrated plate-by-plate using magnitudes
from the Minnesota Automated Plate Scanner POSS-I catalog
(APS\footnote{The APS databases are supported by the National Science
Foundation, the National Aeronautics and Space Administration, and the
University of Minnesota, and are available at
\url{http://aps.umn.edu/}.}, Pennington et al.\ 1993) which are more uniform
than the APM magnitudes because they were calibrated on a
plate-by-plate basis.  The APS catalog was created by scanning the same
POSS-I plate material as the APM and so should be fully consistent with
the APM since it has the same bandpasses, epoch, and so on.

An alternative approach would have been simply to use the APS catalog
in place of the APM catalog to get optical counterparts for the {\it FIRST}
sources.  We preferred the APM catalog because the APS catalog is not
complete over the POSS-I area and does not cover the southern sky at
all, and because the APS catalog retains only sources that appear on
both the red and blue plates, discarding a significant fraction ($\sim$40\%)
of the faint radio source counterparts near the plate limits.
Analysis of the reliability of these single band detections is discussed
in section 5.2 and shows that 92.7\% of the blue-only matches 
within 1" are real, and 97.5\% of the red-only matches within 1" are real
matches.

We matched the entire {\it FIRST} catalog against the APS catalog, extracting
all objects within $20^{\prime\prime}$ of each radio position, and then
matched the resulting list of optical sources with the equivalent
APM/{\it FIRST} match list.  For each APM plate, we determined independent
linear fits (zero-point and slope) for $E$ and $O$ that transform the APM
magnitudes to the APS scale.  The fit was based on APM/APS matches
closer than $5^{\prime\prime}$ that are classified as stellar by the 
APS (which
uses a different photometry method for non-stellar objects.)

For 8 of the 148 POSS-I plates covering the {\it FIRST} area, there were
insufficient APS sources available to determine the photometric
calibration because the corresponding plate was unavailable in the APS
catalog.  For those plates, we bootstrapped a photometric solution
using APM sources in the overlapping regions of neighboring plates.
The set of zero-points and slopes for all 148 plates is available on
the {\it FIRST} website.

\begin{figure*}
\plottwo{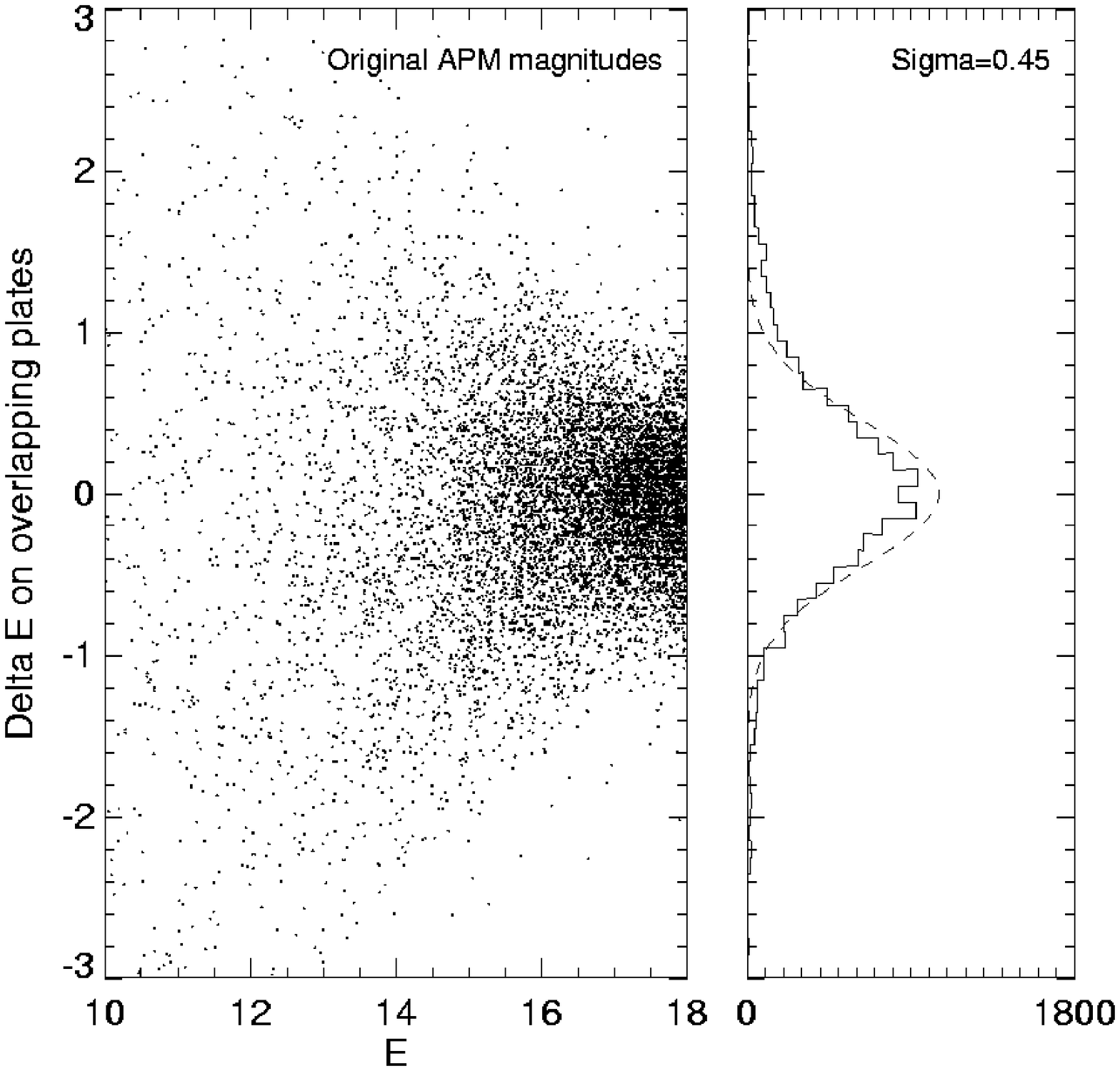}{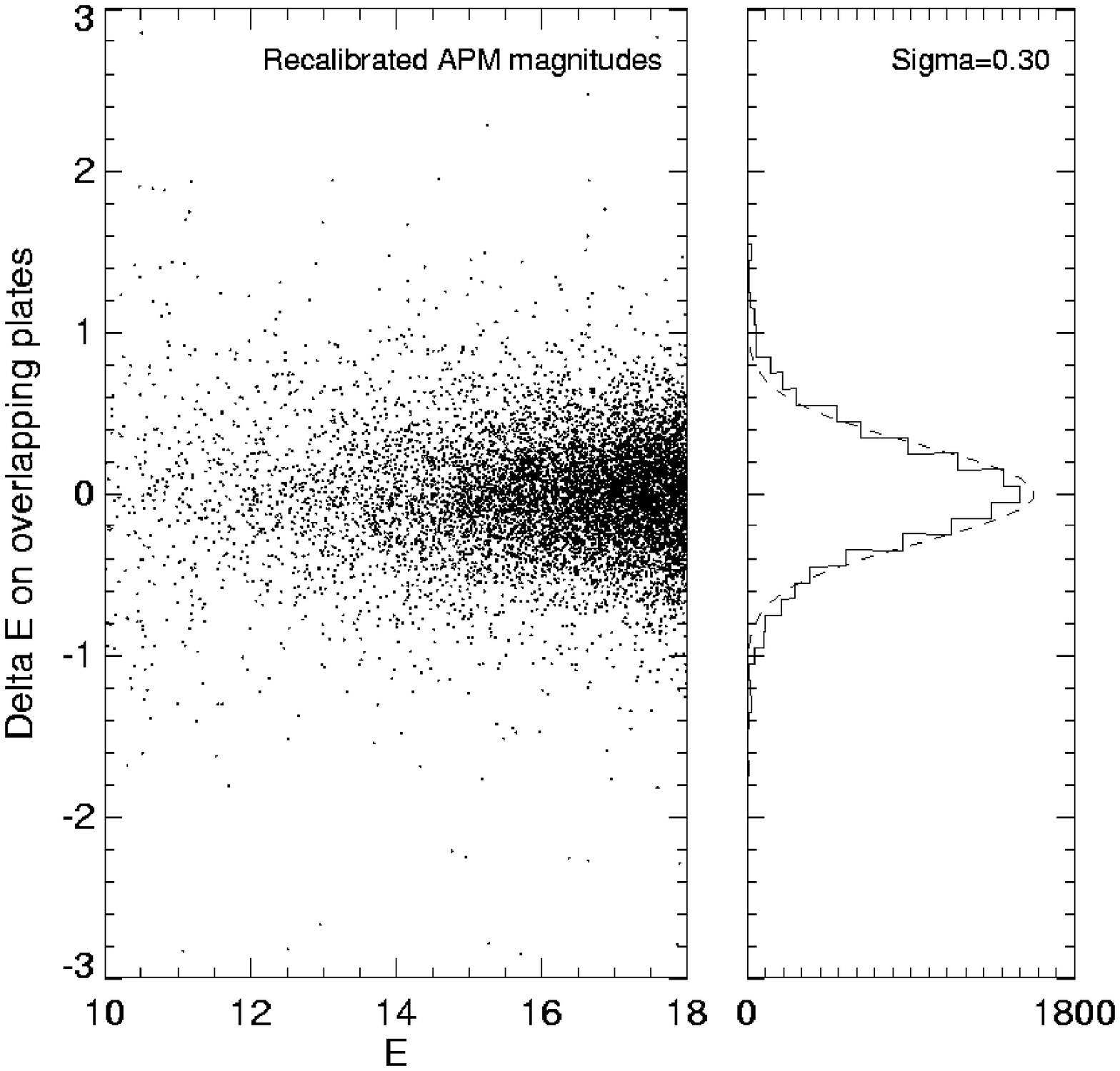}
\caption{
The accuracy of APM magnitudes as determined from a comparison of stellar
objects taken from two different plates in the narrow overlap regions near the
POSS-I plate edges. (a) Magnitude differences using the original APM
magnitudes.  The dashed line is a Gaussian with $\sigma=0.45$.  (b)
Magnitude differences after photometric calibration using the APS
catalog.  The dashed line is a Gaussian with $\sigma=0.30$.  The
recalibrated magnitudes are substantially improved.
}
\label{fig-overlap}
\end{figure*}

This calibration procedure substantially improves the APM photometry.
This is clearly seen in the magnitude discrepancies for APM objects in
the plate overlap regions, which provide two or more independent
magnitude measurements per object. 
Figure~\ref{fig-overlap} 
shows the distribution of
APM magnitude differences before and after calibration; the rms scatter
decreases from 0.45 to 0.30 magnitudes, and the scatter for bright
sources is reduced by an even larger factor.  Since the plate overlap
regions lie at the extreme edges of the POSS-I plates, they are
probably the worst-calibrated areas on the plates; consequently, we
estimate that the recalibrated APM magnitudes are accurate to better
than $\sim$~0.2 magnitudes rms.

Figure 2 shows that the principal problem with the APM magnitudes is
a magnitude-dependent error. We display the differences between the
mean calibrated $O$ and $E$ magnitudes and the original APM magnitudes as
a function of magnitude. Both colors display a quasi-linear trend, with
the total error varying by $\sim0.5$ magnitudes over an eight-magnitude span.
While these curves have been derived from a small subset of all APM
scans, it is likely that application of the corrections they imply will improve
significantly the photometric accuracy of the catalog.

\begin{figure*}
\plottwo{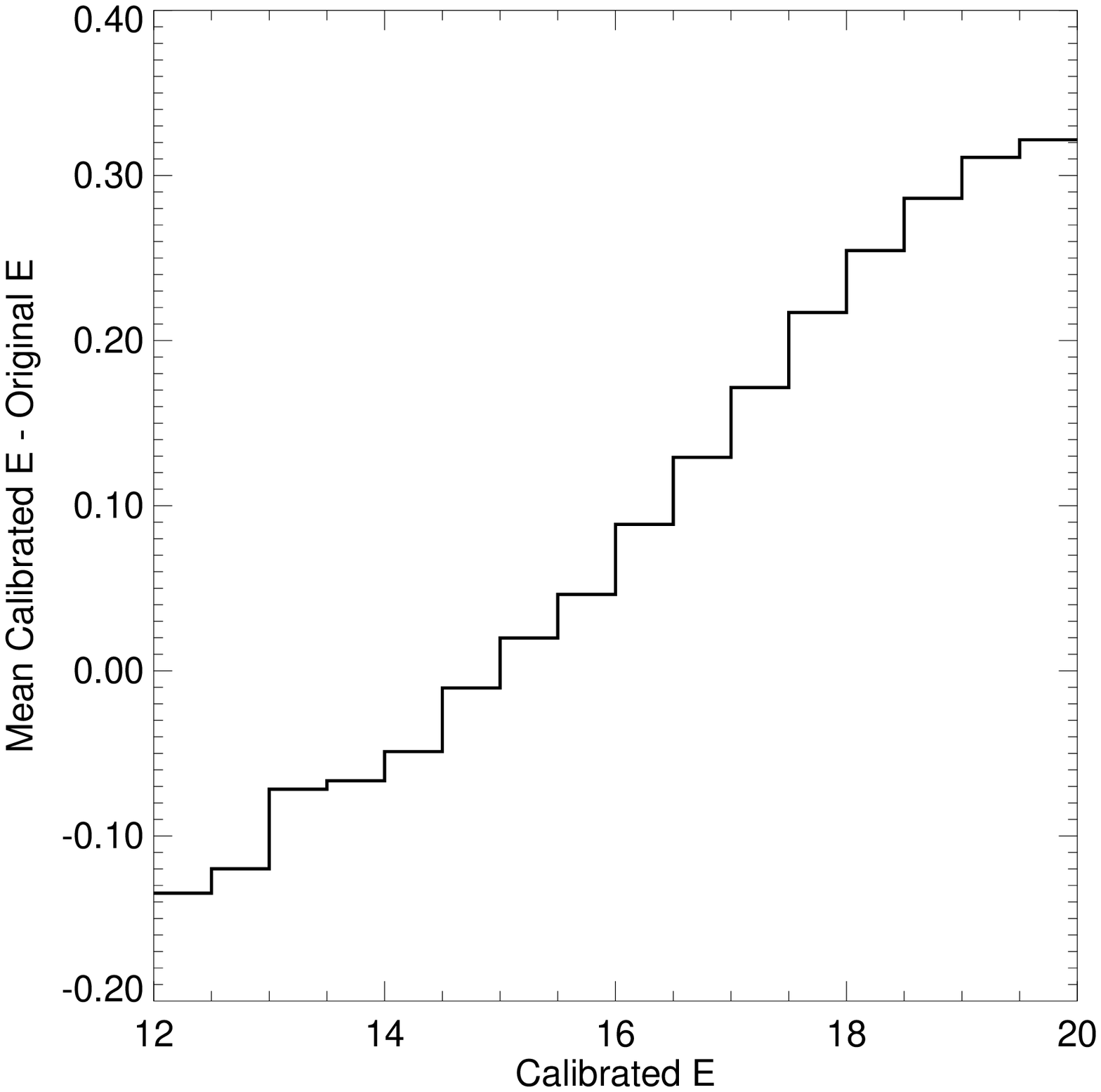}{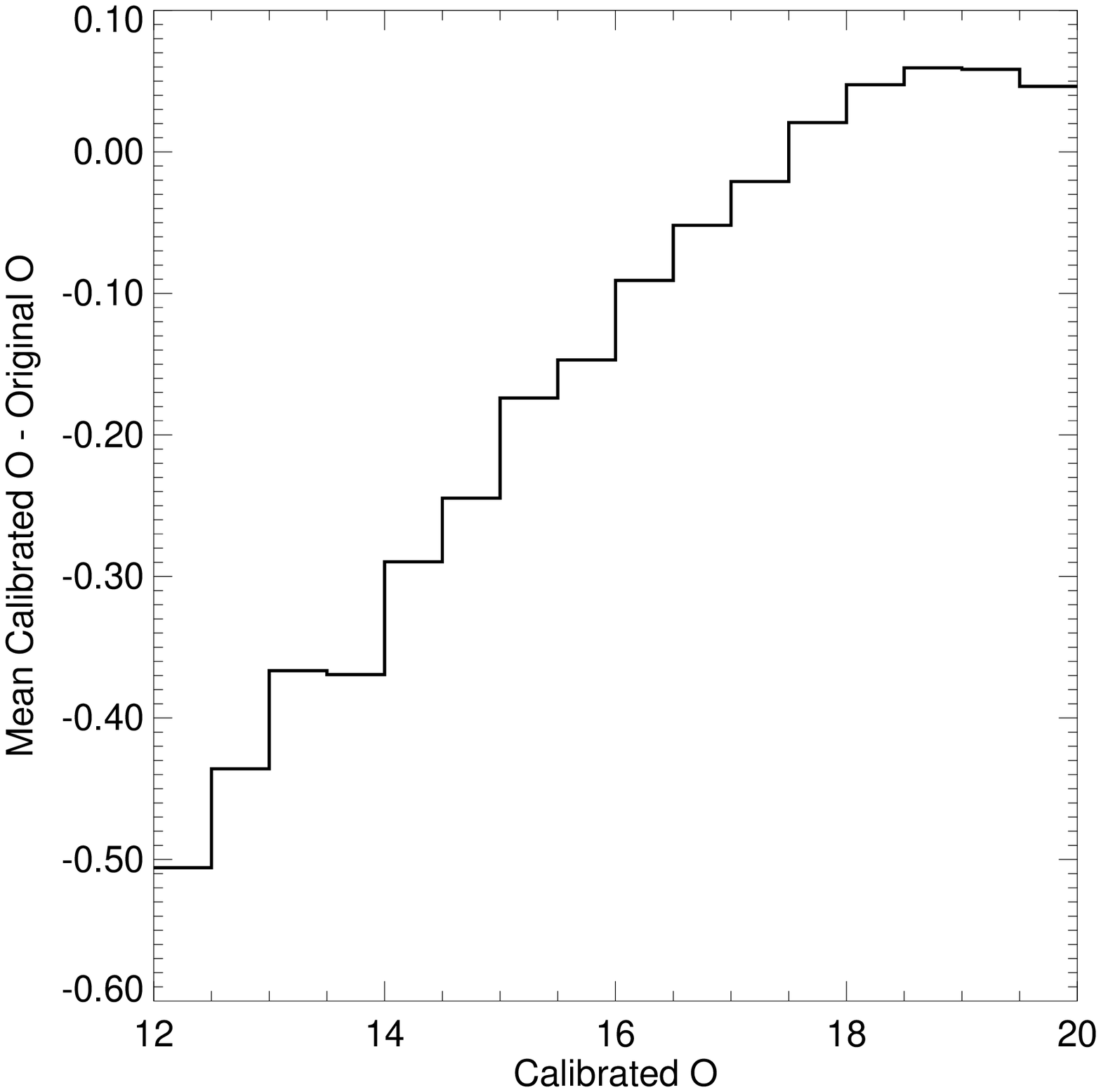}
\caption{The differences between the calibrated and raw APM magnitudes
as a function of magnitude for the (a) $E$ and (b) $O$ plates derived from the
148 POSS-I plates covering the {\it FIRST} survey area.
}
\label{fig-magchange}
\end{figure*}

\subsection{APM Classification}

The APM scans result in a parameterization of each detected image which
includes an $x,y$ position, a peak intensity, a total isophotal
intensity, second moments of the intensity distribution, and areal
profiles (defined as the number of pixels above preset levels which
increase by powers of two above the threshold level). In addition,
a classification parameter is calculated which reports by how many sigma the
object differs from the stellar point-spread function on each plate.
The stellar psf is derived as a function of magnitude to take
into account saturation effects.

These parameters are then used to classify all images into one of four
categories: stellar (consistent with the magnitude- and
position-dependent point spread function, cl$=-1$), non-stellar (a
measurably extended source, cl=1), merged objects (sources with two
local maxima within a single set of connected above-threshold pixels,
cl=2), and noise (objects with nonphysical morphologies, cl=0). 

For
further details of the principles involved, see Maddox et al.
(1991a,b).  Very bright images can often be misclassified, since the
limited set of parameters does not provide an adequate description and
the background-following algorithm attempts to track over them in
order to detect the faint images in the source halos. The
merged/non-stellar boundary is not as reliable as the
stellar/non-stellar boundary, so merged stars are often found in the
non-stellar list (with a smaller number of galaxies in the merged
list). Some objects classified as noise are real; objects found on
both plates are the obvious examples.  Objects classified as noise
which match {\it FIRST} sources are also likely to be real, and we do
not generally exclude these from our analysis.

Bright objects ({\it e.g.}, $O, E < 13$) cover a large number of pixels in the
APM scans and, as a consequence, magnitude and source-size estimates
are very sensitive to small uncertainties in the plate sky level and
details of the background-following algorithm; as a result, large
uncertainties in the parameter estimations can result, and very bright
sources can even be completely missing from the catalog. In addition, bright
galaxies with complex surface brightness distributions can be broken up
into a swarm of discrete sources. At fainter magnitudes, the limitations of
the plate material make reliable separation of stellar and non-stellar
sources problematic. Since our goal here is to determine the
completeness and reliability for the catalog of radio source
counterparts, we do not attempt a comprehensive analysis of the APM
catalog's classification accuracy. Instead we perform several
straightforward comparisons with existing catalogs of bright stars and
bright galaxies and with CCD images which serve to characterize the APM
completeness and reliability with respect to {\it FIRST}
identifications.

\subsection{APM Reliability at Bright Magnitudes}

\subsubsection{Galaxies}

We have examined the DSS images, APM $O$- and $E$-plate catalogs, and {\it
FIRST} images in the vicinity of 1000 entries in the UGC galaxy
compilation (Nilson 1973).  The UGC purports to be an angular-size
limited sample for objects with blue diameters $>1^{\prime}$; it is
complete for $V<14.5$, but includes entries down to $V=18$. This
provides a direct measure of the completeness of our identification of
{\it FIRST} sources with bright galaxies and the accuracy with which
the galaxy parameters are reported in the APM catalog.

Of the 1000 UGC catalog entries, seven were found to have no galaxy
brighter than $V=17$ within $3^{\prime}$ on the DSS. Another dozen
entries had only very small-diameter objects ($<20^{\prime\prime}$) in
the vicinity, while three were extended, but very low surface brightness
objects. Finally, in seventeen cases, there were two (and in one case
three) galaxies of comparable brightness within 1--2$^{\prime}$ of the
UGC position, and it was not possible without further investigation to
distinguish which was the cataloged object.

In only 1 of the 1000 cases was no object present in the APM catalog,
and that case corresponded to one of the extremely low-surface
brightness galaxies; in virtually all cases, in fact, the UGC galaxy was
detected on both plates. Over 275 of the 1000 UGC galaxies are
detected in the {\it FIRST} catalog, and all of these correspond to
objects in the APM source list. Thus, bright galaxies missing from the
APM catalog are not a source of incompleteness in identifying {\it
FIRST} counterparts.

Classifications, colors, and magnitudes for bright APM galaxies are,
however, more problematic. First, for the reasons discussed above,
colors and magnitudes for bright galaxies are at best estimates,
and can be grossly in error for the brightest objects ($E<12$; see
\S3.3). Second, of the 1000 UGC galaxies, 28 were classified as stellar
on both plates; 26 were listed as stellar on the blue plate while
correctly classified on the red, while 82 were classified as stellar on
the red plate, but correctly identified as extended on the blue plate.
Among the {\it FIRST}-detected galaxies, stellar misclassifications had
a higher frequency by a factor of 2 to 3 ($\sim$~20\% in total),
presumably because many of these radio detections represent galaxies
with a bright active (stellar) nucleus.

For $\sim80$ galaxies, entries appear in both the $O$ and $E$ catalogs, but
their centroids are sufficiently far apart that they are not identified
as the same object, and so remain listed as separate objects (without measured
colors) in the merged catalog. These are slightly over-represented
among the radio detections, most likely as a result of the high
radio-detected fraction of interacting/merging galaxies whose complicated
surface brightness profiles confuse the source-finding algorithm.
Finally, somewhat less than 4\% of the UGC galaxies have size and shape
parameters significantly discrepant from the images, either as a
consequence of including a nearby star in the profile, breaking up a
large galaxy into many components, or some other such error; these are
also more common (6\%) among the radio detections owing to the high
fraction of interacting systems {\it FIRST} detects.

In summary, then, the APM catalog is $>99$\% complete for bright
galaxies in the sense that it contains at least one entry at the galaxy
location, and in $\sim97\%$ of the cases, the object is classified as a
galaxy on at least one of the two plates. The descriptions of the
magnitudes, shapes, and colors of the optical objects, however, are
subject to large errors, and should be checked by examination of the
DSS and/or other catalog resources such as NED before they are used for
purposes other than simple identifications.

\subsubsection{Stars}

As noted in the Introduction, the number of stars detected at
centimeter wavelengths is small, and the number of quasars brighter
than $E=14$ is even smaller. Thus, the completeness of the APM for bright
stellar objects is not a major issue in the identification of {\it
FIRST} counterparts.  Nonetheless, for the record, we briefly comment
on APM reliability and completeness for bright stellar objects.

Helfand et al. (1999) report on the detection of 21 radio stars in the
{\it FIRST} region under discussion here with magnitudes in the range
$1.6<V<14.1$.  All but the three brightest objects ($V<4.5$) are
recorded in the APM catalog, and more than half are classified as
stellar on at least one plate. Since most true stellar identifications with
radio sources are impossible without proper motion information,
however, and the number of extragalactic bright stellar counterparts is
vanishing small, any incompleteness at bright magnitudes for stellar
objects in the APM catalog is largely irrelevant to {\it FIRST} source
identification.

\subsection{APM Reliability at Fainter Magnitudes}

In order to assess the completeness and reliability of the APM
counterparts list in the magnitude range $15<E<20$, we use statistics
from work in progress to identify {\it FIRST} sources in the 16 deg$^2$
deep I-band imaging survey ($I<24$) of Postman et al. (1998b; see
Helfand et al. 1998 for a preliminary report on this {\it FIRST}
identification program).

A total of 345 counterparts to {\it FIRST} sources discovered in the
I-band CCD data were selected for comparison with the APM catalog; this
included all 323 objects with $I<20$ and the additional 22 stellar
objects with $20<I<21$. All 123
objects with $I<18.0$ were detected on both plates and over 90\% (75\%)
are correctly classified on the $E$ ($O$) plates. Over 93\% of the 87
objects with $18<I<19$ were also detected on both plates, although the
classifier performs somewhat less well, with 63\% correctly classified
on the $E$ plate. For $I>19$, more than 40\% of the $I$-band objects are
still detected on the $E$ and/or $O$ plates, of which half are correctly
classified on the $E$ plate.  This near to the plate limit, it is clearly
difficult to distinguish between stellar and extended objects;
nonetheless, only one of the 27 spectroscopically identified quasars
with $17<I<21$ in this $I$-band field is misclassified as non-stellar on
both plates. Similarly, we have found that fewer than
5\% of the non-Seyfert quasi-stellar objects in the Veron-Cetty and Veron (1998)
catalog with $R<18$ are mistakenly classified as galaxies on both
plates (Gregg et al. 1996).

Using an APM magnitude-limited sample at $E<19.0$, we find that 62\% of galaxies
are classified correctly on both plates. An additional 28\% are classified correctly on the $E$ plate and are at or below the plate limit on the $O$
plate, while another 4\% are classified correctly in $O$ and incorrectly in $E$;
only 6.5\% are misclassified on both plates. For stellar images, the score
is similar, with 14/26 correct on both plates and another seven
classified correctly on one plate. In summary, for $E<19$, 92\% of all objects
are classified correctly on at least one plate.

As a further check on the incompleteness of the APM catalog, we have
compared a subset of the database with the APS catalog of the POSS-I
(Pennington et al.\ 1993). The
APS catalog contains only objects detected on both plates, so it does
not include the faintest APM objects which are often detected on only a single
plate.  This is nonetheless a very useful test of completeness and
accuracy for the APM catalog, since the APS started with the same plate
material but used completely independent scanning hardware and
processing software.

The APS catalog test set contains 61,000 unique objects that match 71,000
different APM objects within a matching radius of 10\arcsec; the objects
were drawn from areas on the plates within $20^{\prime\prime}$ of
{\it FIRST} sources, and thus include both real radio source identifications and
random background sources.  There are
532 APS objects (0.9\%) that have no matching APM source.  Most of
these are very near the catalog limit: 364 have $E>20$ or $O>21$,
leaving only 168 (0.3\%) that are reasonably bright.  Checking the
Digitized Sky Survey reveals that the great majority of these objects
are blended with other nearby objects in the APM catalog (and
usually classified as such), leading to poor agreement in the
positions from the two catalogs. Furthermore, we should
note that we have {\it not} examined the plates by eye at the locations of
these ``missing'' sources, and some could be spurious APS entries which
are legitimately absent in the APM catalog. Thus, the
fraction of objects that is simply missing from the APM catalog above
$E=20$ ($O=21$) is very small, certainly less than 0.1\%.

\begin{figure*}
\epsscale{0.45}
\plotone{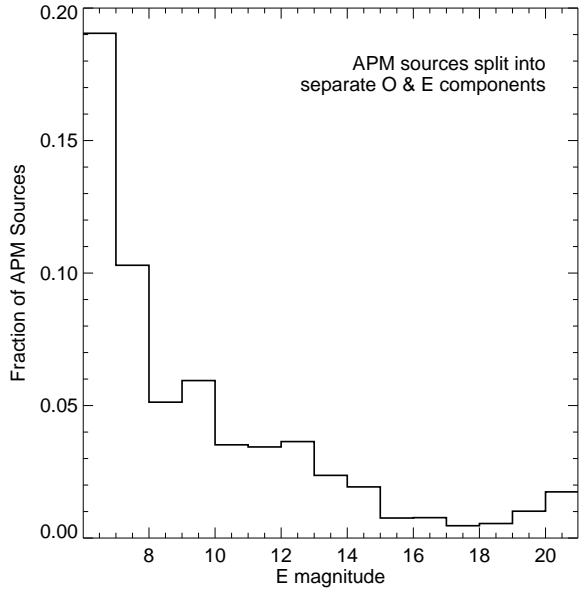}
\epsscale{1}
\caption{
The distribution as a function of APM $E$ magnitude of the fraction of objects
that appear
in the APM catalog as a pair of sources, one detected only on the $O$ plate
and one detected only on the $E$ plate.  These split objects are
quite bright and are typically either large complex galaxies or
close, blended objects. The fraction of such
objects in the range of most {\it FIRST} identifications ($15<E<20$) is
very low.
}
\label{fig-splitmag}
\end{figure*}

Occasionally the positions measured by the APM on the $O$ and $E$ plates
are sufficiently different that the red and blue detections are not
recognized as being a single source and so two entries (one $E$-only and
one $O$-only) appear in the APM catalog (\S 3.6.1).  This can lead to incomplete
radio source identification.  In our APS test sample, there are
$\sim730$ cases (1.2\%) where a single APS source matches a close pair
($<10$\arcsec) of $E$-only and $O$-only objects from the same APM plate.
We consider these objects to be cases where a single source has
been split into two catalog entries.  The majority of these split
objects are bright (Fig.~\ref{fig-splitmag}) and are further examples of
the complex and/or blended objects that occasionally caused trouble for
the bright UGC galaxies discussed above. The median $E$ magnitude of
these objects is 16.1; the fraction of such objects is $<1\%$ for the magnitude
range 15--20 in which most radio counterparts are found.

\section{FIRST as an Astrometric Standard}
\label{sectionastrometry}

\subsection{Introduction}

As we show in WBHG, all unresolved {\it FIRST} sources down to the
survey limit have 90\% confidence positional uncertainties of
$<1\arcsec$; furthermore, the absolute reference frame is tied to the
VLB reference frame with an offset of $<0.03\arcsec$ and an rms
uncertainty of $<0.15\arcsec$ in each coordinate. This level of
precision is unprecedented for such a high-surface-density catalog
({\it e.g.}, the Guide Star Catalog entries have an rms uncertainty of
0.4\arcsec\ with systematic errors of up to 1\arcsec), making {\it
FIRST} a useful astrometric calibration standard for any astronomical
catalog which contains a  significant number of {\it FIRST} source
counterparts. In this section, we use the radio source positions to
calibrate and correct the APM intraplate astrometry errors and the
plate-to-plate shifts, allowing an improvement by nearly an order of
magnitude in POSS-I positions.

The {\it FIRST} catalog contains about 90 sources deg$^{-2}$ of which
$\sim$15\% (see \S 5.2) have APM optical counterparts within
$2^{\prime\prime}$, yielding $\sim$600 matches per POSS-I plate. This,
in effect, provides a dense grid of astrometric standards across each
plate.  More important, the optical counterparts to {\it FIRST}
sources are generally faint, and thus complement the astrometric
calibration that can be achieved using bright standards such as the
PPM and Tycho catalogs that have been used heretofore as the basis of
the APM plate solutions.

\subsection{Overall APM Astrometric Offset}

Whilst working on radio identifications for the Jodrell Bank-VLA
Astrometric Survey (JVAS -- Patnaik et al. 1992;
Hook et al. 1996; Snellen et al. 2001), it was
noted that there was evidence for a shift between the mean Right
Ascension and Declination of the APM reference frame and the VLA
reference frame.  The mean shift was derived by taking the median of
the nearest optical counterparts to the survey's flat-spectrum radio
emitters and yielded:

\begin{itemize}
\item RA$_{APM}$(corrected) = RA$_{APM}$ (original) + 0.35 $\pm 0.05$ arcsec
\item Dec$_{APM}$(corrected) = Dec$_{APM}$ (original) + 0.35 $\pm 0.05$ arcsec
\end{itemize}

We verified that this effect was not intrinsic to the PPM reference frame
used for the APM plate solutions at that time by comparing
the optical positions of a set of VLBI Radio Reference Frame objects
(Johnston et al. 1995) with optical counterparts on APM scans and found
that the same systematic shift was present. We then remeasured POSS
field 1393 centered on the North Galactic Pole in two orientations: with the normal
scanning direction and with the plate rotated by 180 degrees.  We found
that, after application of the original astrometric alignment using the
PPM stars, the derived positions for faint ($>14^{th}$ magnitude)
objects differed by $\sim0.7^{\prime\prime}$ in both coordinates.  This
indicated that the APM measurement system was introducing a systematic
shift in the positions of bright objects with respect to fainter ones.
The origin of this effect is still under investigation. Our first step
in comparing {\it FIRST} sources and the APM catalog, then,  was to
take out this systematic shift of 0.35\arcsec\ in both coordinates.

\subsection{Intraplate Errors}

\begin{figure*}
\plottwo{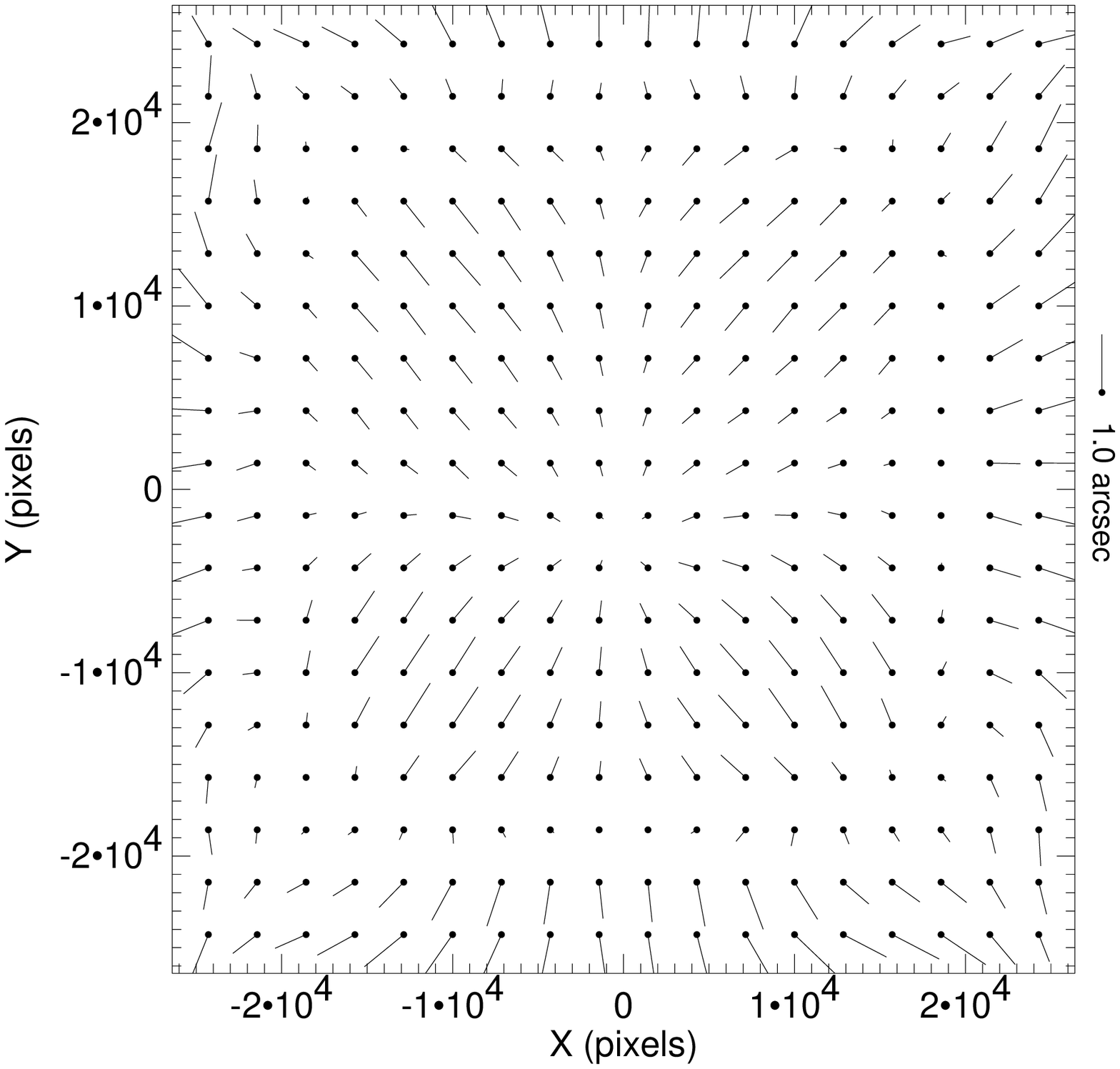}{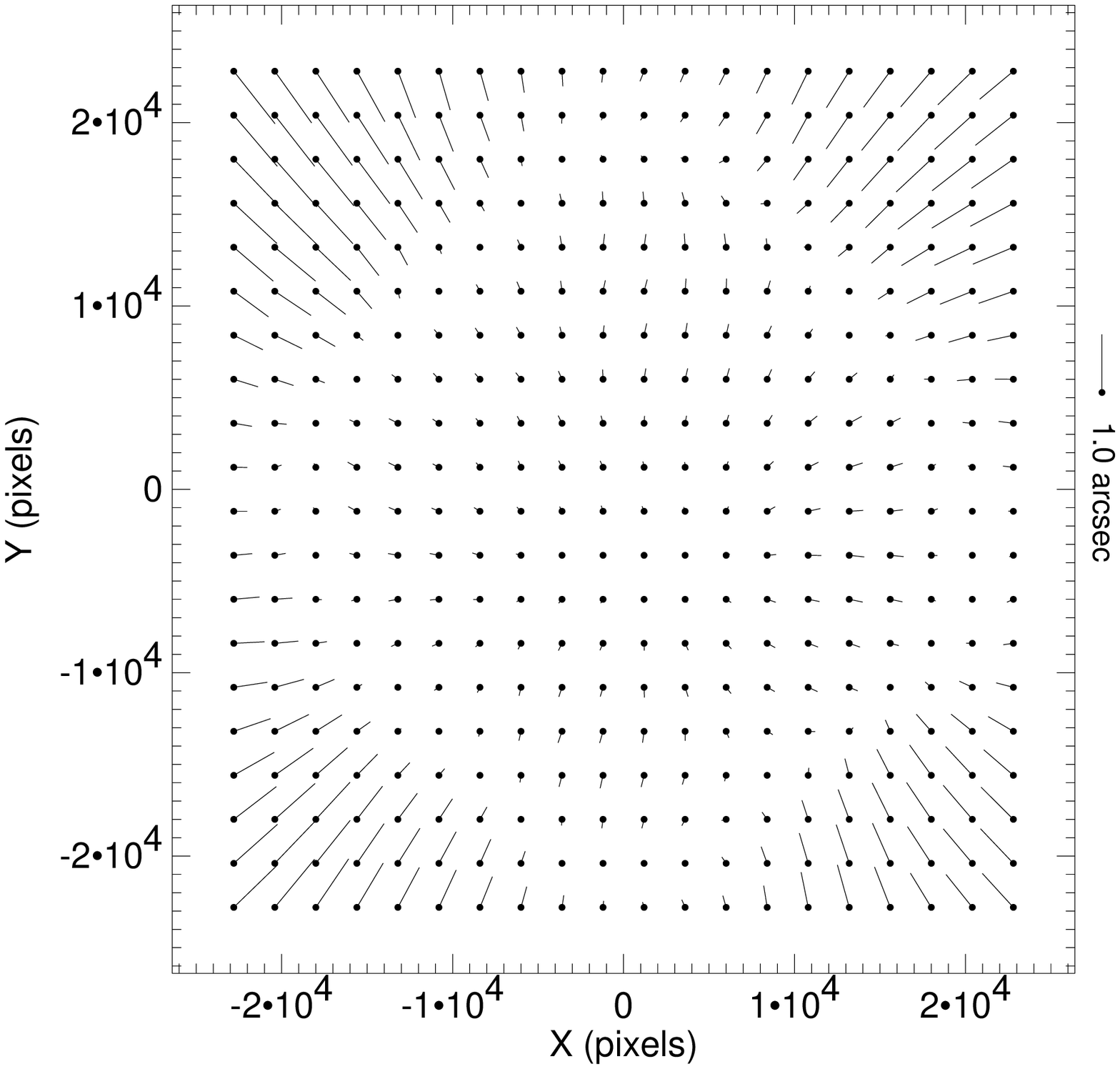}
\caption{
(a) The distortion in raw APM catalog positions as a function of position on POSS-I
plates derived by comparison to reference positions from the TYCHO
catalog of astrometric reference stars.  The lines show the average
position shifts in $20^{\prime} \times 20^{\prime}$ regions across the plates
(6667 pixels = 1 degree); the scale of the shifts
is shown on the right.  There are large, systematic distortions
that are not modeled by the APM plate solution.
(b) The distortion remaining in the APM positions after correction for
the distortions shown in (a).  These are derived by comparing corrected
APM positions to {\it FIRST} positions for APM-{\it FIRST} matches.  The distortions
are small near the plate center, but beyond $\sim2.5^\circ$ from plate
center there is a strong, increasing radial distortion.  This is
a magnitude-dependent position error in the APM catalog; the positions
of bright sources (such as TYCHO stars) are properly corrected by application
of the
distortion map in (a), but positions of faint objects (such as the
optical counterparts to {\it FIRST} radio sources) are actually made worse
by the bright-star correction map.
}
\label{fig-mask}
\end{figure*}

The original POSS-I plates suffer from various distortions resulting
from the stress induced by the plate holder, the vignetting of the
48-inch Schmidt telescope, etc., which are reproducible from plate to
plate (Irwin 1994).  In Figure~\ref{fig-mask}(a), we display the
pattern of residuals from the TYCHO reference stars derived by stacking
the APM-TYCHO offsets for all such stars on 148 plates.  Systematic
errors reach 1.3\arcsec\ in some regions of the plate. A correction for
this effect has, as noted above, been applied to all APM catalog
positions using a correction map derived from the Tycho catalog.

\begin{figure*}
\epsscale{0.45}
\plotone{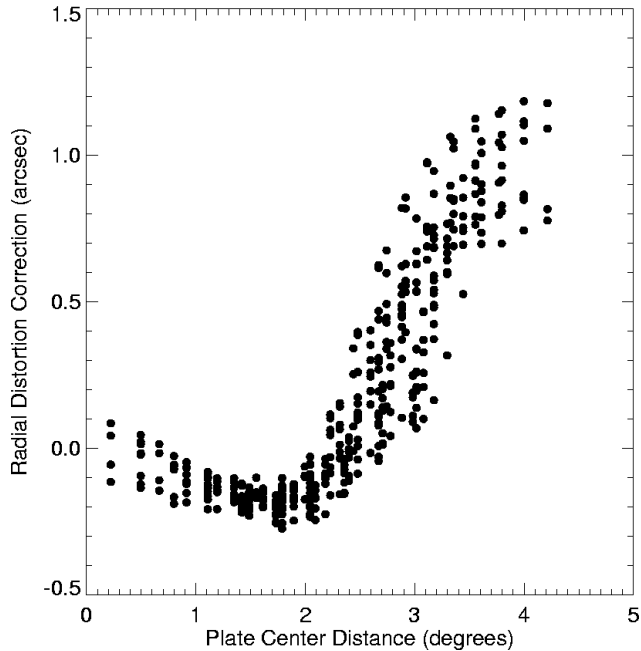}
\epsscale{1}
\caption{
The radial offset between APM and {\it FIRST} positions as a function of
distance from plate center; each point represents one of the $20^{\prime}
\times 20^{\prime}$ grid cells into which we have divided the plates.
}
\label{fig-radial}
\end{figure*}

We compared these corrected APM positions with those of {\it FIRST}
radio sources over the 148 POSS-I plates which subsume the current
survey region.  Including all sources with optical counterparts having
offsets $\le2.0\arcsec$ ($\sim$60,000 objects with a false rate of
$\sim5$\% --- see \S 5) yields the result displayed in
Figure~\ref{fig-mask}(b).  Residuals within a radius of 2.5 deg of the
plate centers are substantially reduced compared with Fig.~\ref{fig-mask}(a),
but large radial errors with
magnitudes up to 1.2\arcsec\ are now seen around the plate
edges (Fig.~5). We believe this effect is a consequence of a combination of
telescope vignetting and plate saturation of the bright stars in the
Tycho grid: vignetting distorts the stellar images in the radial
direction, producing elliptical isophotes whose centroids are
systematically shifted toward the plate center. The {\it FIRST}
counterparts, largely much fainter objects for which
saturation is not a problem, reveal this distortion.

To correct for this effect, we have chosen to calibrate the errors
using all {\it FIRST} counterparts with $15.0 < E < 19.5$; the bright
limit is chosen such that the magnitude-dependent errors described
above will be unimportant (and only excludes 6\% of the matches),
while the faint limit is imposed in order to include only those objects
for which the random errors are small. The correction proceeds
iteratively. We calculate a mean offset (in both the radial and
tangential directions) for all sources in the stacked image in cells
$20^{\prime}$ on a side (20$\times$20 cells cover the plate). We then
apply the derived offsets to all APM objects in the catalog based on
their $x$-$y$ plate positions, rematch to {\it FIRST}, and then
recalculate the mean offset for the cell. The process is repeated until
it converges ($\sim$4 steps).  An iterative approach is required
because the APM position shifts are comparable to the $2\arcsec$
APM-{\it FIRST} matching radius.  The resulting APM-{\it FIRST}
position errors are much improved:  the maximum residual mean offsets
are less than 0.18\arcsec\ with an rms of only 0.06\arcsec\ over the
whole plate.

This intraplate correction is strictly valid only for the range of
magnitudes used in the calibration ($15.0 < E < 19.5$). An examination
of the offsets for fainter counterparts suggests that the corrections work
well to the plate limit, although the positional uncertainties increase
slightly owing to the statistical fluctuations which are inevitable in
deriving positions for objects near the detection threshold. For the
brighter objects, however, the derived positions of candidate
counterparts to {\it FIRST} sources suffer from the same type of
distortion evident in the Tycho-based solution, presumably in a
magnitude-dependent way.  Rather than attempting to derive the
functional form of the magnitude dependence from the small number of
bright {\it FIRST} counterparts, we recommend that for very bright
objects either the separation for an acceptable match be increased
or the offset be computed using 
the APM astrometry without the {\it FIRST} astrometric correction.
For example, in the {\it FIRST} Bright Quasar Survey (White
et al. 2000) we
include all sources with optical matches closer than $1.2^{\prime\prime}$ in
{\it either} astrometric frame, a procedure which adds 24 objects to
the 1214 selected using the {\it FIRST} astrometric solution alone.
Since the typical discrepancy between the two solutions is
$<1\arcsec$ and our minimum matching criterion is $\ge 1\arcsec$, no
genuine matches will be missed if this procedure is followed. However, since
the overwhelming majority of optical counterparts are faint, we have used only the
matches with the {\it FIRST} corrections applied when compiling the match
statistics in this paper.

\begin{figure*}
\epsscale{0.45}
\plotone{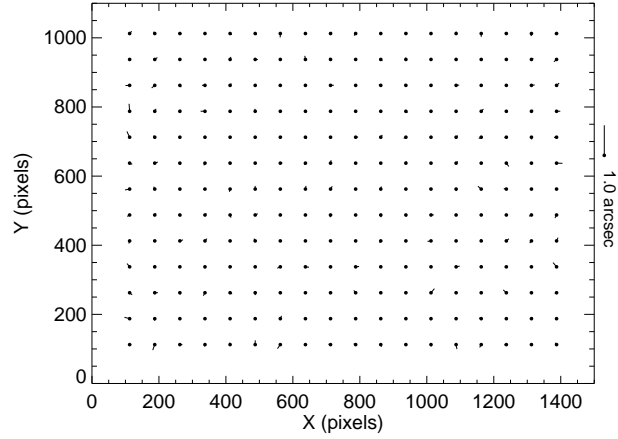}
\epsscale{1}
\caption{
The distortion in {\it FIRST} catalog positions as a function of position within
the {\it FIRST} coadded images.  The scale of the vectors is shown on the
right; 1 pixel = $1.8^{\prime\prime}$.
No systematic errors in the {\it FIRST} positions
are evident, indicating that we have successfully corrected the
geometric distortions in the {\it FIRST} images.
}
\label{fig-fpmask}
\end{figure*}

This use of {\it FIRST} sources as astrometric calibrators assumes,
of course, that there are no systematic errors in the radio positions.
In Figure~\ref{fig-fpmask}, we display the analogous plot to
Figure~\ref{fig-mask}(b) for the {\it FIRST} catalog: the $x$-$y$
positions of each {\it FIRST} source in the radio ``plate'' ({\it i.e.},
coadded image --- see BWH) are extracted and the mean offset from the
APM matches to these sources (binned in $2^{\prime}$ cells) is
calculated. The uniformity is excellent and the magnitude of the errors
is very small:  they are all $<0.13\arcsec$ with an rms of
0.05\arcsec\ over the whole field\footnote{We display only the region
within the 1150$\times$1550 pixel image from which sources actually
enter the catalog; sources closer to the boundaries will be extracted
from the adjacent coadded images.}.  This result, coupled with our
global astrometric calibration described in WBHG suggests our
confidence in {\it FIRST} astrometry is justified, and that our radio
catalog can be utilized routinely to aid the astrometric calibration of
other catalogs.

\subsection{Plate-to-plate Shifts}

\begin{figure*}
\epsscale{0.45}
\plotone{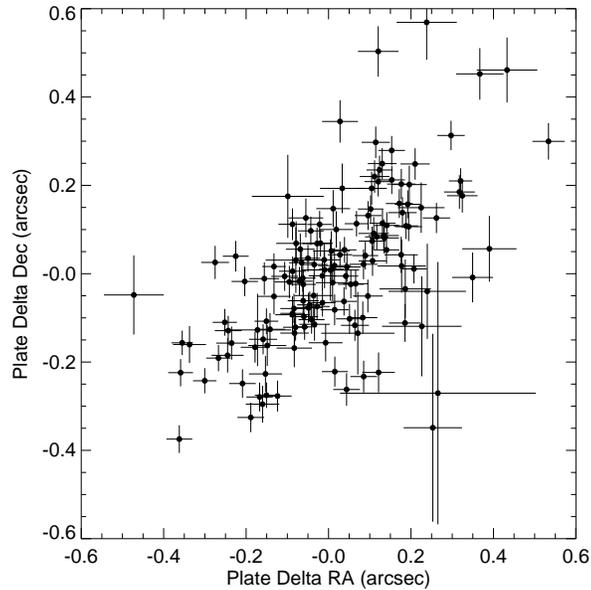}
\epsscale{1}
\caption{
The mean shifts for each APM plate in the {\it FIRST} survey area derived
by comparing APM positions (corrected for the intraplate distortions in
Fig.~4) to {\it FIRST} positions.  Each point shows the RA and Dec shifts
for a plate with $1\sigma$ error bars.
}
\label{fig-plate}
\end{figure*}

Having established the global APM offset and calibrated out the
reproducible intraplate errors, we are left only with translations of
individual plates with respect to our radio reference frame. The mean
{\it FIRST}-APM offsets for each of the 148 plates in the survey region are
shown in Figure~\ref{fig-plate}.  These offsets were computed using an
iterative approach similar to that used to derive the plate distortion.
There is no obvious trend with Right Ascension or Declination, but
shifts remain at levels ranging from 0.01\arcsec\ to
0.70\arcsec; the distribution of errors is roughly Gaussian in both
coordinates with an rms of $\sim 0.16^{\prime\prime}$ and an apparent
tendency for large offsets in one coordinate to be matched by those in the
other. We remove these shifts by simply subtracting the shift
computed for a plate from the APM positions of all objects on the
plate. Note that for plates which fall completely within the boundaries
of the {\it FIRST} survey, this calibration is final; however, for
those plates only partially covered by the existing survey, the global
offset will be redetermined as new data accumulate. The size of any
subsequent correction is expected to be no more than $\sim$0.1\arcsec. The
current values, along with the intraplate correction matrix are available
on the {\it FIRST} Website.

After correcting for the plate shifts, we have recomputed the
plate-based distortion maps (Fig.~\ref{fig-mask}) to determine whether
the systematic plate shifts affect them.  The changes in the
distortions are small ($<0.1\arcsec$).

Having completed the astrometric calibration, we can now proceed to
generate a catalog of optical identifications for the {\it FIRST}
survey.

\section{Optical Identification of FIRST Radio Sources}
\label{sectionids}

The primary condition for assigning an optical counterpart to a radio
source has traditionally been positional coincidence; while other
factors can enhance or detract from the probability that the
association is correct, these are generally second-order effects. If
the positional accuracies for both the radio and optical catalogs are
very high, positional coincidence alone can afford a high degree of
confidence. In BWH, we calculated the expected error rate for
associations between the APM and several hypothetical VLA radio surveys
with differing angular resolutions. The reliability of associations
based on positional coincidence were presented as a function of optical
source density on the sky, and as a function of apparent V magnitude at
the North Galactic Pole.  The results demonstrated the strong dependence of both
reliability and completeness on the angular resolution of the
radio survey.

In reality, the actual error rate in radio/optical associations is a
complex problem depending on the details of the relative astrometric
accuracy of the two data sets, the optical morphology of candidate
objects, and radio source morphology. In this section we quantify
empirically the reliability of associations between {\it FIRST} radio
sources and optical counterparts on the POSS-I plates as measured by
the APM.

\subsection{Background Rates Using Closest Matches}

In our analysis of the {\it FIRST}-APM match rate, we retain only the closest
optical match to each radio source and vice versa.  While the optical catalog
we present below includes all matches out to $20^{\prime\prime}$,
not just the closest one,
using only the closest matches when describing the identification statistics
makes understanding the background and true match rates
easier.  We want to know the number of radio sources that have optical
counterparts, but we want to avoid double-counting ({\it e.g.}, a particular
optical object may be the closest counterpart for two different radio
sources.)  Also, neither catalog can have pairs of sources with separations of
less than a few arcseconds, which means that there is effectively a
``hole'' in the catalog around each radio or optical source.  This
makes it very difficult to calculate chance coincidence rates, since
true matches suppress additional false matches in a complicated
manner.

We can avoid both of these problems by using only the closest matches.
This also modifies the chance background rate, since once a source has
a counterpart there can be no additional counterparts at wider
separations.  This effect is straightforward to understand
and incorporate into the background computation: the area being searched
for new matches at a given radius is contributed only by {\it FIRST} sources
that do not already have a closer optical counterpart.  Another
complication when using closest matches is that the matching circles
for close pairs of radio sources can overlap.  An optical source that
falls in the overlap region between two radio sources can only be
assigned as the counterpart to one of them.  This is also a
mathematically tractable problem.  Both effects are treated as a
reduction in the effective sky area that is searched as a function of
matching separation.

When both of these effects are taken into account, it is possible to
predict the number of chance (false) coincidences as a function of the
radio-optical separation.  A simple model for this prediction that
assumes a uniform, random distribution of optical sources on the sky
predicts more coincidences at large separations than are actually
seen, owing to the variation in optical source density
$\rho_{opt}$ on the sky.  In the presence of a varying optical source
density, radio sources that remain unmatched at large separations are
more likely to be found in low-density areas of the sky; consequently,
the effective mean optical source density decreases as the separation
increases.  If the variation $\delta\rho_{opt}/\rho_{opt}$ is not too
large, this can also be modeled fairly simply.  For an ensemble of $N$
radio sources, the effective background source density is
\begin{equation}
\rho_{eff}(r) = \bar{\rho}_{opt} - \sigma_{opt}^2 \left( \sum_i A_i(r) \phi_i(r) \over
\sum_i \phi_i(r)\right) \quad,
\end{equation}
where $\bar{\rho}_{opt}$ is the mean source density, $\sigma_{opt}$ is
the rms variation in the density, $A_i(r)$ is the area already searched
for source $i$, and $\phi_i(r)$ is the angular part of the matching
circle being searched for source $i$.  For an isolated {\it FIRST} source
with no nearby neighbors (so there are no overlapping error circles)
and an optical match at radius $r_0$, we have
\begin{eqnarray}
\phi_i & = & 2\pi \quad , \quad r < r_0 \quad  \\
       & = & 0 \quad , \quad r > r_0 \quad ,
\end{eqnarray}
and
\begin{eqnarray}
A_i & = & \pi r^2 \quad , \quad r < r_0 \quad  \\
       & = & \pi r_0^2 \quad , \quad r > r_0 \quad .
\end{eqnarray}
For ``gregarious'' {\it FIRST} sources in close groups, the matching circles overlap.
Then $\phi_i$ will include only the portion of the error circle at $r$ that
would not be assigned to another {\it FIRST} source, so that in general
$\phi_i \le 2\pi$.

Our background model, then, has two parameters: the mean
source density $\bar{\rho}_{opt}$ and the variance in the density
$\sigma_{opt}^2$.  Values for these quantities were estimated from a spurious
match catalog generated by offsetting the coordinates of the radio
sources by $5^{\prime}$ to the south.  We use this procedure
instead of simply adopting the match rate between, say, 10\arcsec\ and
20\arcsec\ around the real source positions in order to minimize any
enhancement in the false rate resulting from the presence of real
radio-optical associations at large separations.  Such associations
result both from optical counterparts to multiple-component {\it FIRST}
objects (where the optical position need not match any one of the
cataloged {\it FIRST} components closely) and clustering of galaxies
around {\it FIRST} objects (which will produce an uncharacteristically
high optical source surface densities in the vicinity of the radio
sources.)  The parameters for the background due to various types of
APM objects are given in Table~\ref{table-background}.

\subsection{Match Results}

\begin{figure*}
\epsscale{0.45}
\plotone{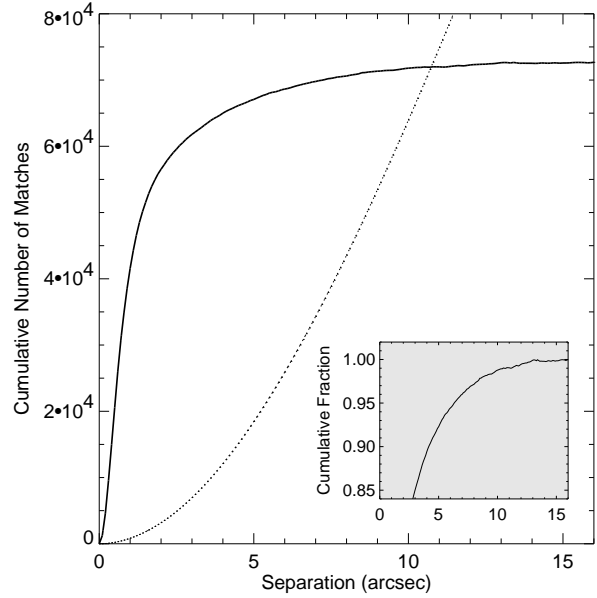}
\epsscale{1}
\caption{
The cumulative number of {\it FIRST}-APM matches as a function of
separation.  Only the closest APM object to each {\it FIRST} source is
included; both isolated and ``gregarious'' {\it FIRST} sources (those with
near neighbors) are included.  The background coincidence rate (dotted line)
has been subtracted.  The inset shows the cumulative fraction of
sources as a function of radius with an expanded scale.  The vast
majority of matches within $2^{\prime\prime}$ are clearly true identifications;
a few true identifications are found out to $>10^{\prime\prime}$ separations.
A total of $\sim73,000$ {\it FIRST} sources have optical counterparts in the APM
catalog.
}
\label{fig-cumnumber}
\end{figure*}

In Figure~\ref{fig-cumnumber}, we display the result of matching all
382,892 radio sources in our catalog to the astrometrically corrected
APM catalog. The plot shows the cumulative excess of matched sources
over the background of chance coincidences as a function of the offset
between the radio and optical positions.  We estimate that 98\% of the
APM sources within 1\arcsec\ of a {\it FIRST} source are physically
associated with the radio source; 42,400 sources meet this criterion.
Even out to 2\arcsec, 94.5\% of the 59,700 associations are real.
Figure~\ref{fig-cumnumber} indicates that some real matches occur out
to $>10$\arcsec\ although the reliability decreases steadily as the
separation increases. Integrating under the curve of
Figure~\ref{fig-cumnumber} out to 4\arcsec\ implies 61,800 real
associations (16\% of all radio catalog entries). Inside the 1\arcsec\ radius,
16\% of the optical counterparts are classified as stellar on both
plates, while 41\% are classified as non-stellar on both plates; in the
remaining cases, the classifiers disagree or the object is only
detected on one plate.  Note that, at faint magnitudes, classification becomes difficult
owing to the limited number of pixels above threshold; in addition, active
nuclei can make a galaxy appear stellar in the blue band, and the bulges of
faint ellipticals and SOs are generally unresolved. Thus, galaxies are
increasingly classified as stellar as the plate limit is approached.

In Table~3
we tabulate the completeness and reliability of objects that
are only detected on a single red or blue plate. These have been estimated
by shifting the radio positions by 3~arcmin in declination and running the
matching analysis in an identical manner on this shifted dataset.
This shows that  92.7\% of the blue-only matches 
within 1" are real, and 97.5\% of the red-only matches within 1" are real
matches. This is a lower limit on whether these single band
detections are real objects since we expect some of the the
chance associations to be real celestial objects.

The matching results are more complicated to analyze when we take into account
the complex radio sources that get broken into two or more components in
the {\it FIRST} catalog.  We call these ``gregarious'' sources (as
distinguished from isolated sources) because they are found in clumps
on the sky. Note that some gregarious sources are simply the chance
superposition of two unrelated radio emitters at different distances;
nonetheless, from our vantage point, they are gregarious, and confuse
the matching statistics in a manner similar to that of the real multi-component
objects. We define the sociological boundary between gregarious and isolated
at $60^{\prime\prime}$; {\it i.e.}, any source with no other catalog entry within a radius of $60^{\prime\prime}$ is classified as isolated, and all
other sources are labeled gregarious. Such a fixed boundary is arbitrary, and,
indeed, a small number of very extended objects will be incorrectly classified
as isolated; we discuss this matter further in \S5.
 
\begin{figure*}
\epsscale{0.45}
\plotone{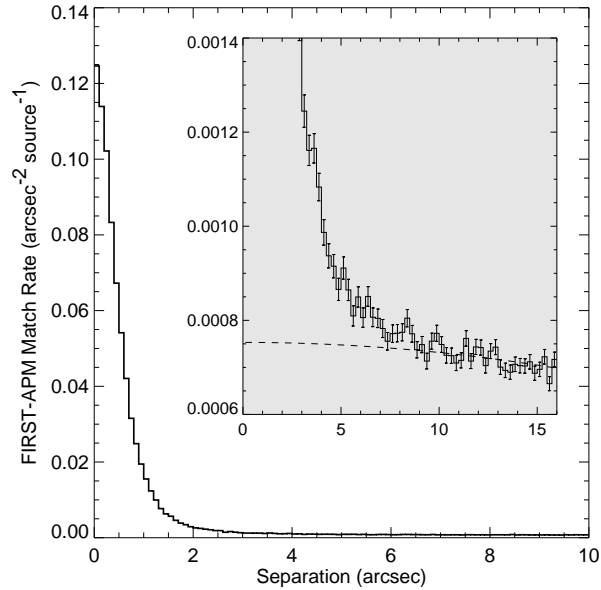}
\epsscale{1}
\caption{
A histogram of the number of matches as a function of
separation between the {\it FIRST} and APM positions.  Only the
closest APM match to each {\it FIRST} source is included, and only
isolated {\it FIRST} sources have been included to avoid the
ambiguities in the background computation for multiple-component
{\it FIRST} matches (see Fig.~\ref{fig-sephist_double}.)
The histogram
has been normalized by the product of the area of each annular bin and the
number of {\it FIRST} sources, so it gives the mean number of APM
matches per unit area for each {\it FIRST} source.
The background
rate of coincidental matches
is shown in more detail in the inset (error bars
are $1\sigma$ and the dashed line shows the expected chance
coincidence rate.)
}
\label{fig-sephist}
\end{figure*}

\begin{figure*}
\epsscale{0.45}
\plotone{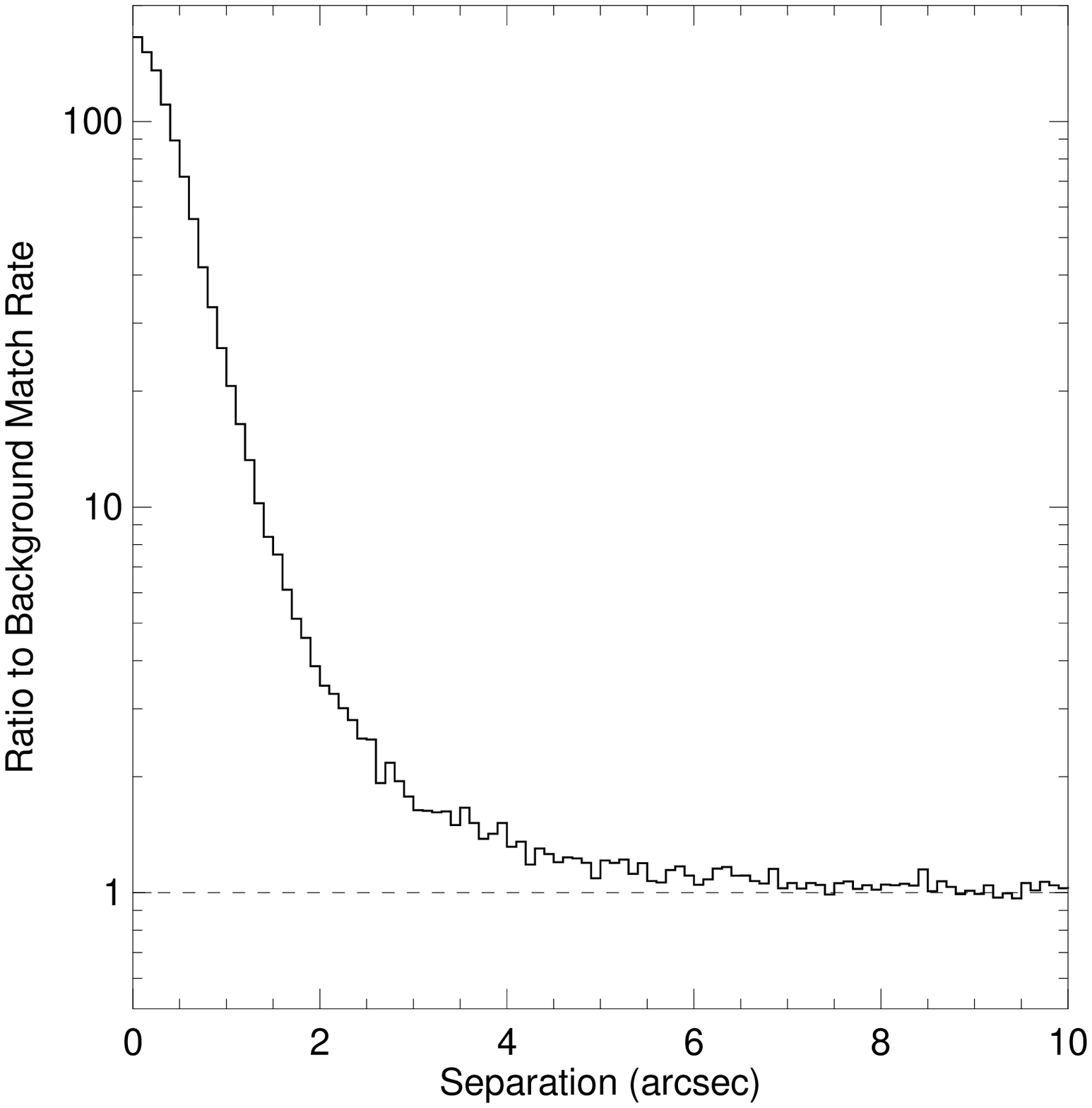}
\epsscale{1}
\caption{
The ratio of the number of APM-{\it FIRST} matches to the expected number of
coincidental matches as a function of separation.  As for Fig.~\ref{fig-sephist},
only the closest APM matches to isolated {\it FIRST} sources are included.
This function
shows the effect of increasing the matching radius between
the catalogs.  At a separation of $\sim2.5^{\prime\prime}$, the
background rate per unit area is approximately equal to
the true match rate, so a small increase in the matching
circle size leads to about equal numbers of true and false
matches being included.  Within $2\arcsec$ the vast majority
of the matches (98\%) are real associations; when non-isolated
{\it FIRST} sources are included, this fraction declines to 95\%.
}
\label{fig-sephist_rat}
\end{figure*}

In Figure~\ref{fig-sephist}, we display the number of isolated matched sources
as a function of the offset between the radio and optical positions in
0.1\arcsec\ bins, normalized by the annular area of each bin.  Figure~\ref{fig-sephist_rat} displays the ratio of
the number of matches to the predicted false rate as a function of
separation.

\begin{figure*}
\plottwo{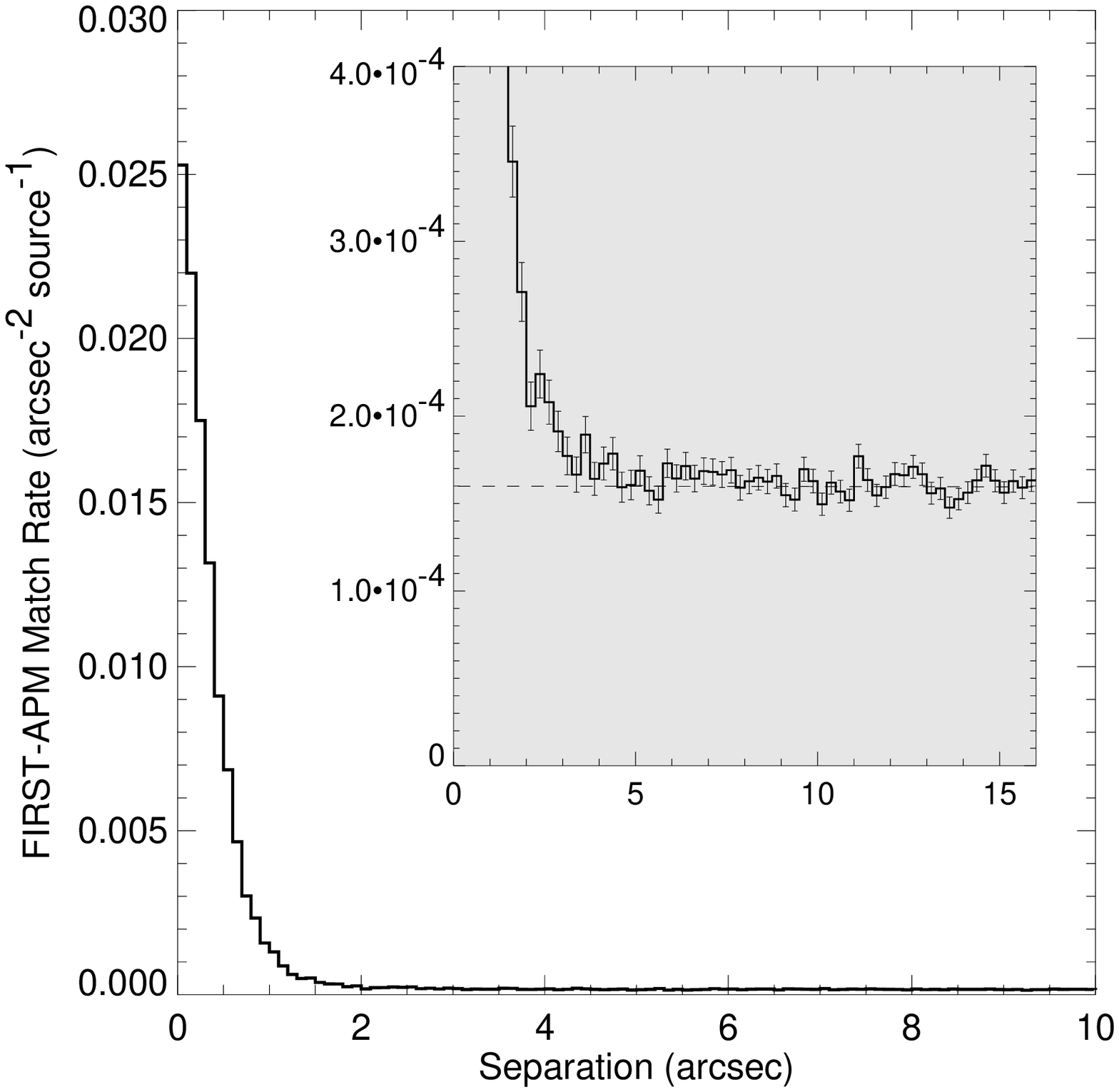}{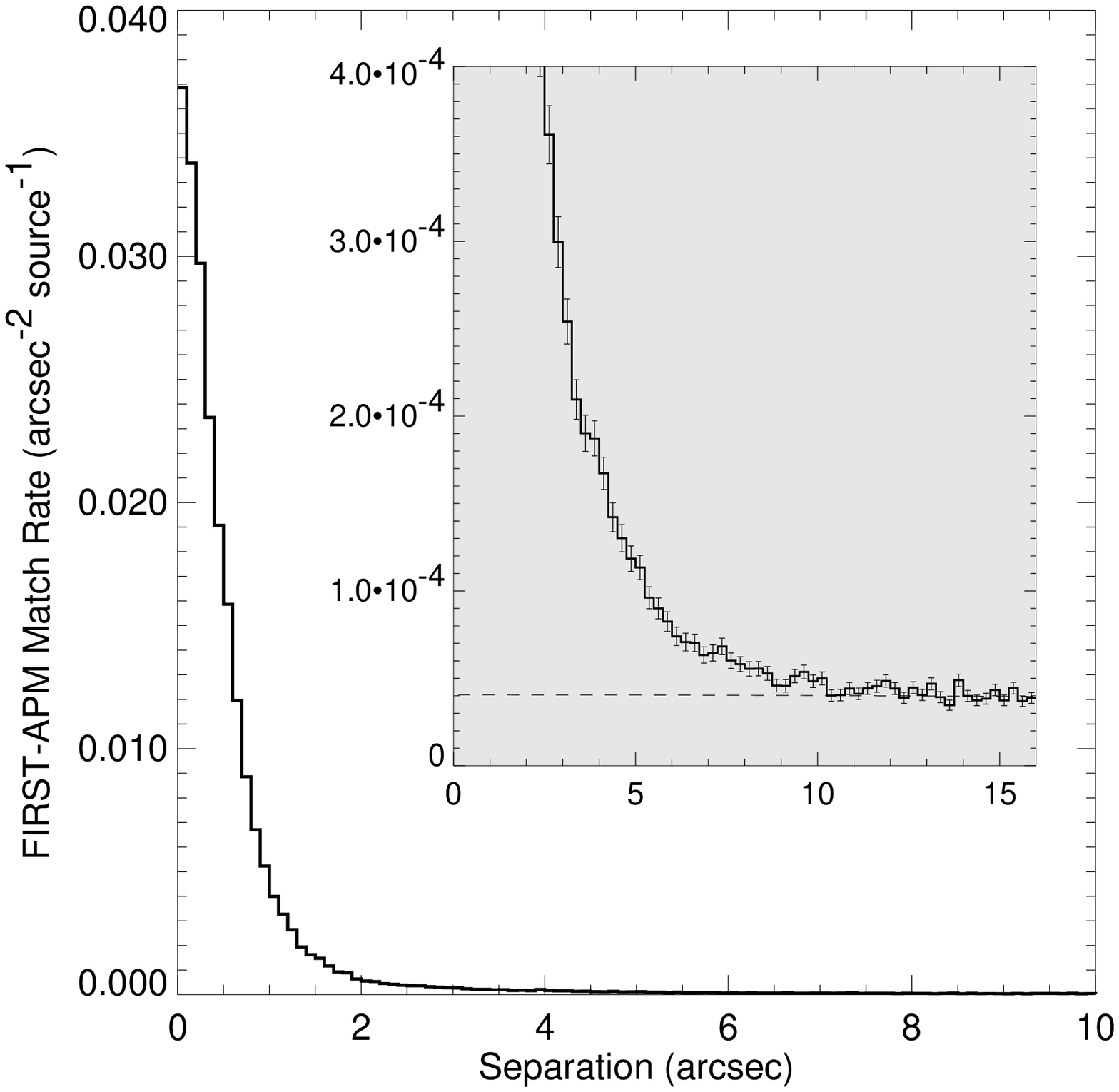}
\caption{
The match rate as a function of separation for (a) APM objects classified
as stellar on both the POSS-I plates and (b) APM objects classified
as non-stellar on both plates.  As for Fig.~\ref{fig-sephist},
only the closest APM matches to isolated {\it FIRST} sources are included,
and the insets show the counts to larger radii with a much
expanded scale.  Galaxies are less concentrated to small separations
than stars (mainly because the optical positions are less well-determined),
but the background rate for galaxies is four times smaller than that for
stars, which makes matches to galaxies reliable to larger radii
than those for stars (see Fig.~\ref{fig-relccomp_star_gal}).
}
\label{fig-dclose_star_gal}
\end{figure*}

Both the background rate and the typical angular separation between
{\it FIRST} and APM positions depend strongly on the optical
classification. Figure~\ref{fig-dclose_star_gal} shows the distribution
of separations for sources classified as stellar on both plates or
non-stellar on both plates. The {\it FIRST}-APM positions typically differ by
more for galaxies, which have less well-determined optical positions.
However, it is easier to identify galaxies confidently as {\it FIRST}
counterparts because the background rate for galaxies is four times
smaller than for stars (see Table~\ref{table-background}).

\begin{figure*}
\plottwo{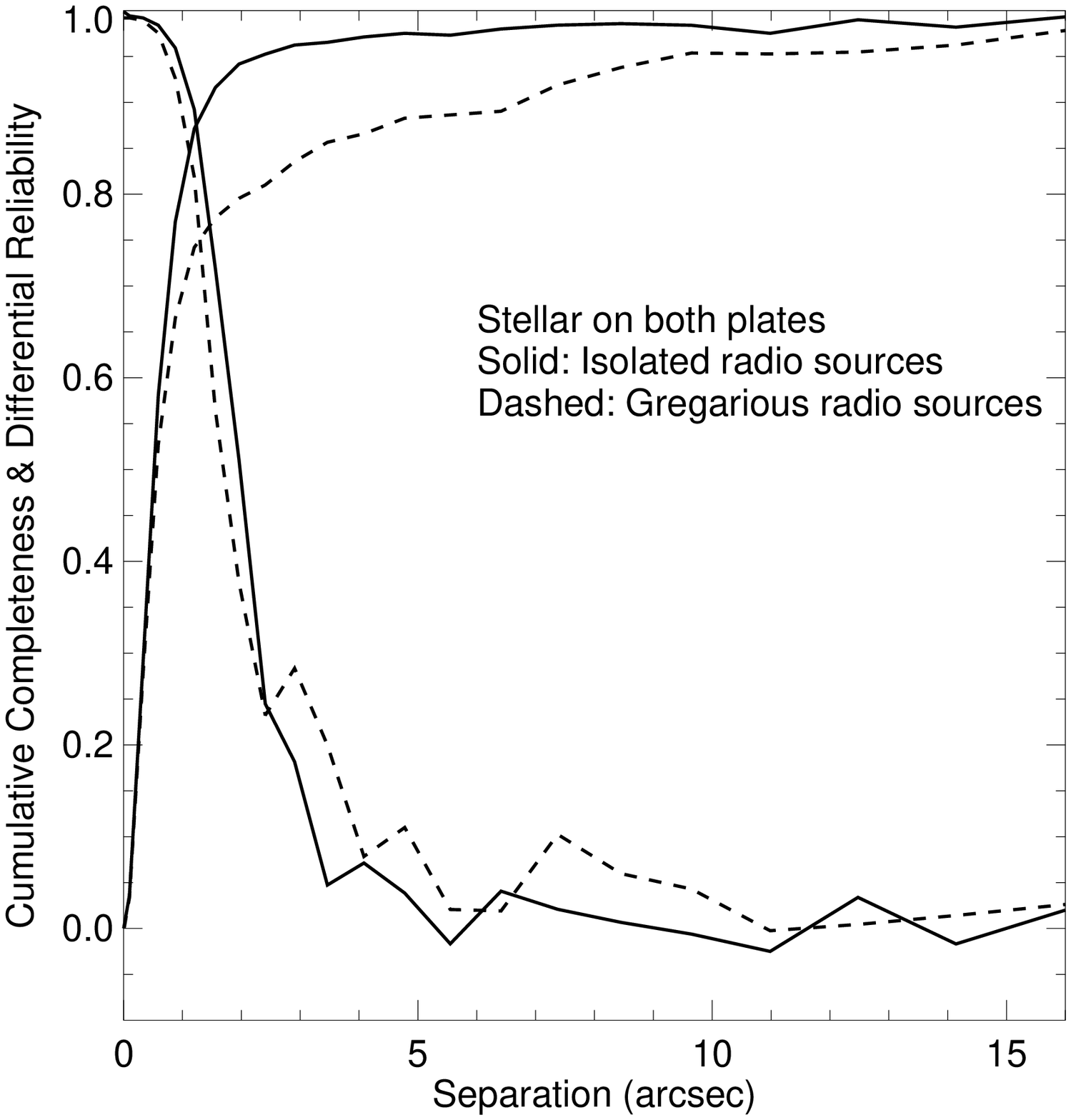}{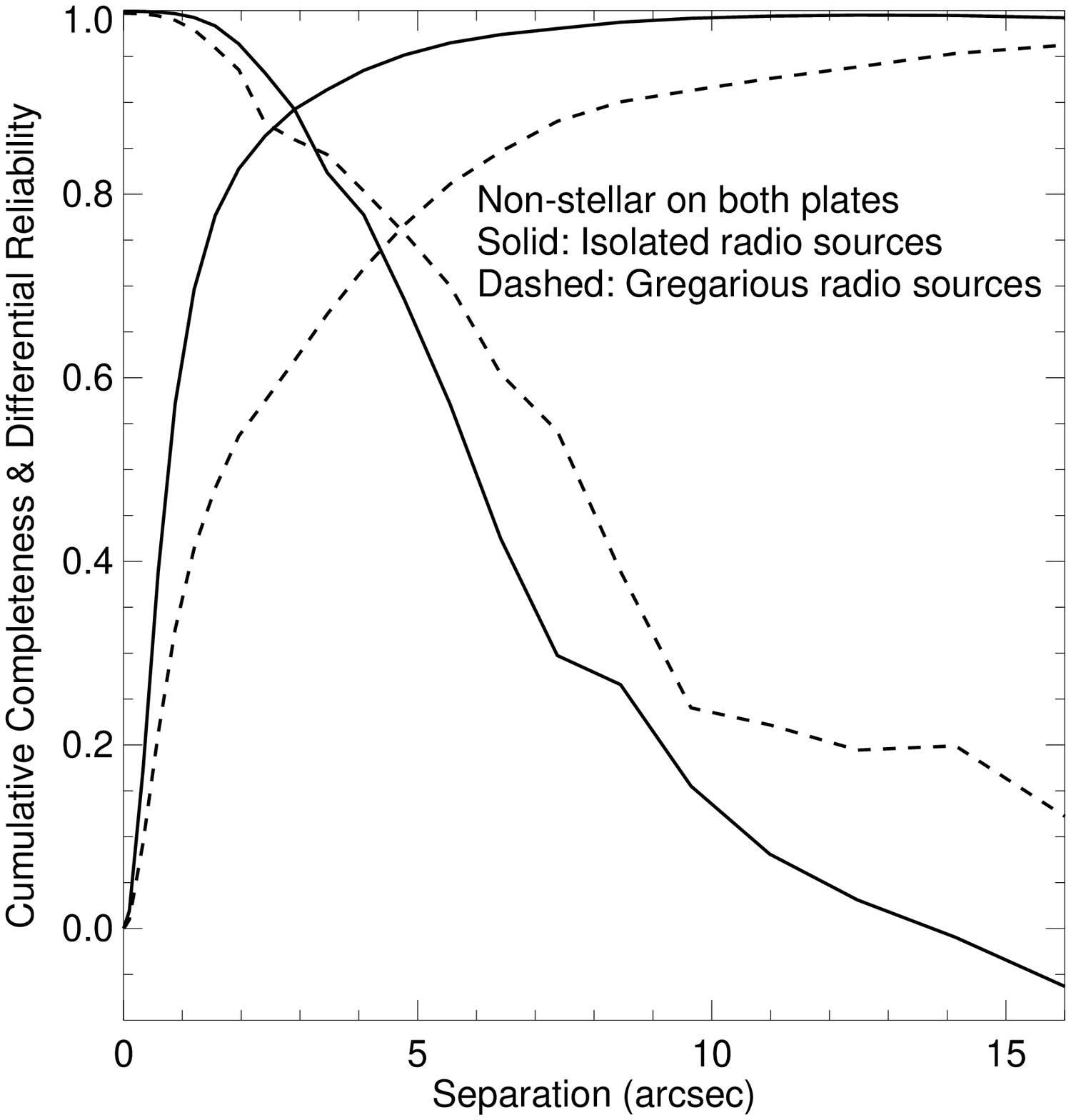}
\caption{
Completeness and reliability as a function of separation for (a)
objects classified as stellar on both APM plates and (b) objects
classified as non-stellar on both plates.  The completeness rises
almost monotonically from zero to one as the separation increases (the
small deviations from monotonicity are due to the background
subtraction).  The differential reliability, which measures the
probability that sources at a given separation are true associations
rather than chance coincidences, declines toward larger separations.
The solid lines show the rates for isolated {\it FIRST} sources, while the
dashed lines are for gregarious {\it FIRST} sources.  Stellar matches to
isolated sources are very closely concentrated toward small
separations, and few stellar matches beyond a few arcseconds are real.
On the other hand, because the background rate for non-stellar matches
is much smaller (see Fig.~\ref{fig-dclose_star_gal}), matches with
galaxies are reliable to much larger radii.
}
\label{fig-relccomp_star_gal}
\end{figure*}

We quantify this effect by calculating the angular radius which
contains 90\% of the real associations as well as the percentage of
false matches within the same radius.  For example, these numbers are
4.2\arcsec\ and 17\% for the distribution of all {\it FIRST}/APM sources as
shown in Figure~\ref{fig-cumnumber}. If we restrict the match to APM
objects that are classified as galaxies on both plates
(Fig.~\ref{fig-dclose_star_gal}b), the values are $4.0^{\prime\prime}$ and
2.7\%.  The 90\% radius is quite similar to that for all associations,
since galaxies constitute the majority of all identifications; however,
the reliability for galaxies is much higher because the background rate
for galaxies is smaller
(Fig.~\ref{fig-dclose_star_gal}).  By comparison, if we restrict
ourselves to APM sources that are stellar on both POSS plates
(Fig.~\ref{fig-dclose_star_gal}a), the 90\% radius and error rate are
$1.7^{\prime\prime}$ and 6\%.  The smaller footprint of stars on the POSS-I leads
to more accurate optical positions (hence the smaller 90\% radius).
These and other cases are summarized in Table~\ref{table-confidence}
and in Figure~\ref{fig-relccomp_star_gal}, which displays the
completeness and reliability of matches as a function of separation.

\subsection{The Effects of Radio Morphology}

The statistical agreement between {\it FIRST} and APM positions is also
affected by the morphology of {\it FIRST} sources.  If we restrict the {\it FIRST}
sample to point-like radio sources (but include all APM sources), the
90\% association radius is 2.0\arcsec\ with a false rate of 5\%.
Limiting the discussion to sources that are point-like in both the
radio and the optical results in 90\% of the matches within 1.1\arcsec\ with a 
2.4\% false rate.

\begin{figure*}
\epsscale{0.45}
\plotone{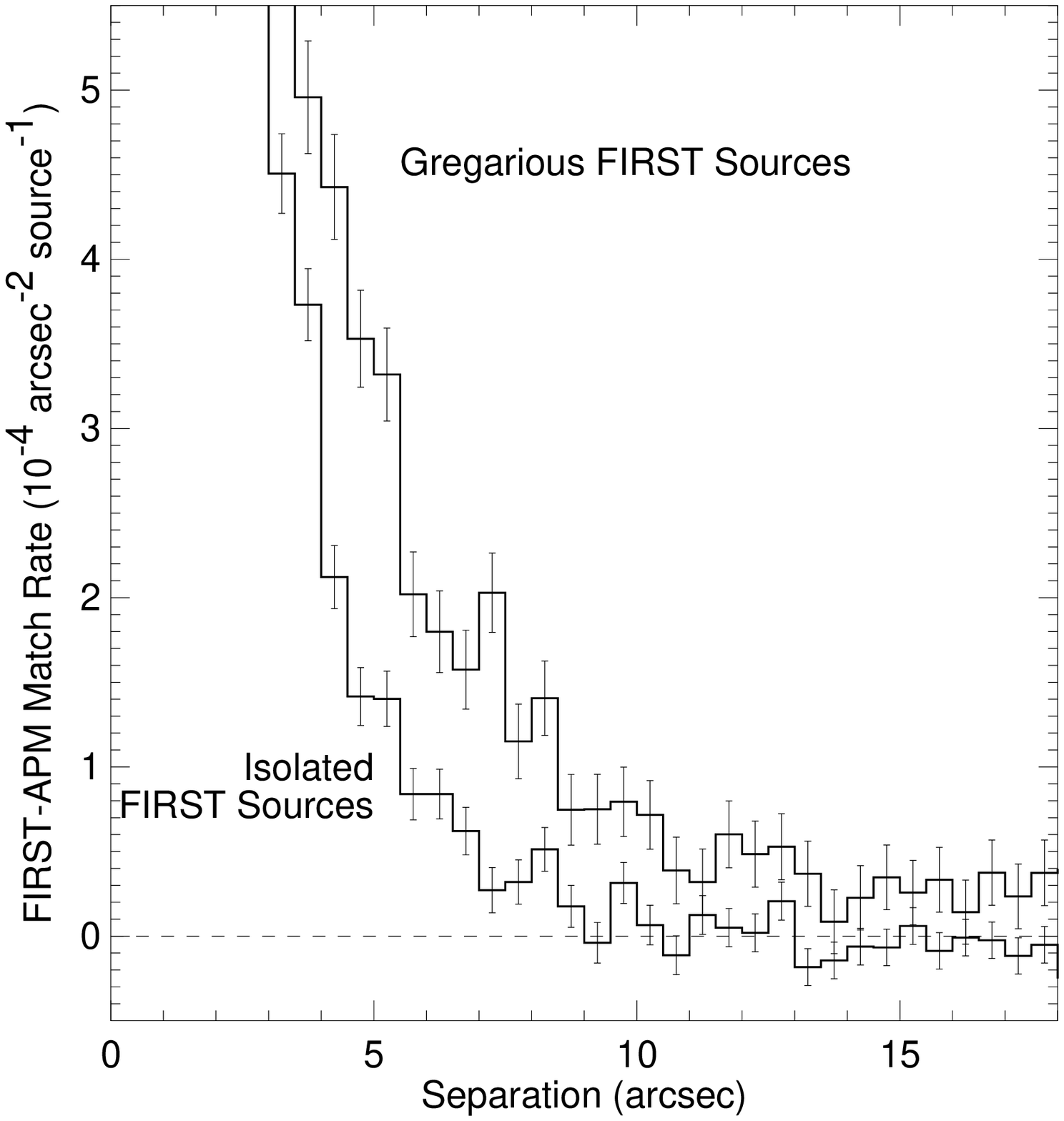}
\epsscale{1}
\caption{
The normalized histogram of separations for isolated {\it FIRST} sources
(with no other cataloged {\it FIRST} objects within $60^{\prime\prime}$) and
``gregarious'' {\it FIRST} sources (which do have neighbors in the
catalog.)  The background coincidence rate has been subtracted.
Clearly most of the matches at large separations
are attributable to the multiple-component gregarious sources, where
the position of the optical counterpart is not as tightly coupled
to the {\it FIRST} radio position as it is for isolated sources.  Developing a
matching strategy for gregarious {\it FIRST} sources and determining the
chance coincidence rate is a difficult problem which we begin addressing in
\S5.4.
}
\label{fig-sephist_double}
\end{figure*}

Figure~\ref{fig-sephist_double} compares the histogram of
separations for isolated sources with that for the gregarious sources.
It can be
seen that gregarious sources contribute most of the matches at large
separations; indeed, they show evidence for a statistically significant
excess of matches even beyond $15^{\prime\prime}$.  Many of the gregarious {\it FIRST}
sources are components of classical double radio sources, which may
have no radio component at the center of the double.  Consequently the
optical counterpart will typically be $\sim1/2$ the double separation
from each component. We now begin to explore counterpart identification
strategies for such sources.

\subsection{Optical Counterparts to Double Radio Sources}

The wide variety of radio source morphologies (not to mention the extent to which
their appearance depends on the resolution of the survey) makes it difficult to
design a robust algorithm that predicts the
locations of the optical counterparts to multiple-component
sources. We present here an empirical approach to the problem for the simplest
class of such sources -- isolated doubles.

\begin{figure*}
\epsscale{0.45}
\plotone{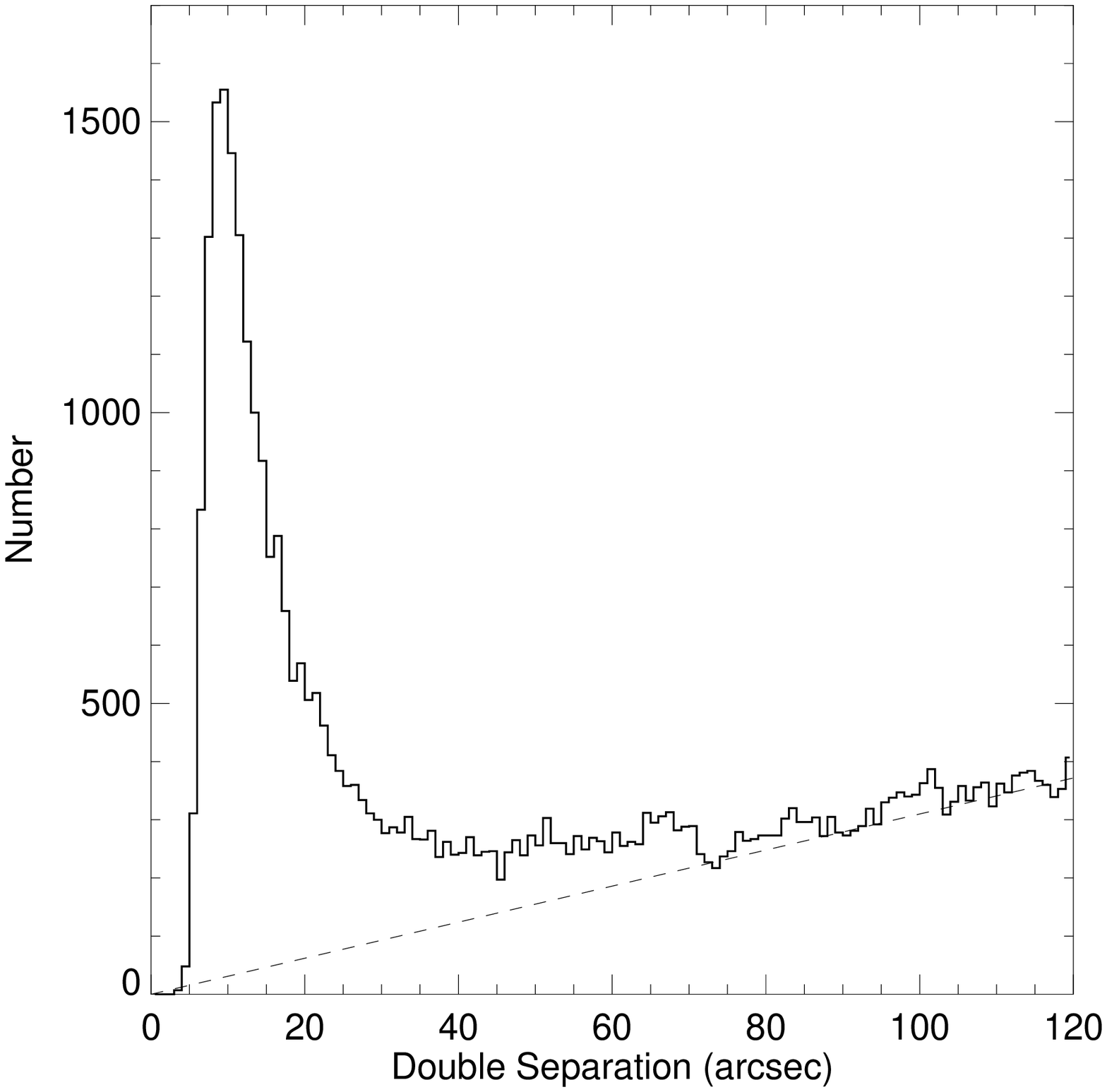}
\epsscale{1}
\caption{
The distribution of angular separations of all isolated pairs
of {\it FIRST} sources (no third component within $120^{\prime\prime}$). The
minimum separation of $3^{\prime\prime}$ is set by the resolution of the
{\it FIRST} images and our detection algorithm;
the fall-off beyond $10^{\prime\prime}$ arises from a variety of effects (see text).
The dashed line represents an estimate of the chance coincidence rate.
At a separation of $40^{\prime\prime}$, approximately half the doubles are
physically associated and half are chance superpositions of unrelated sources
}
\label{fig-dblangsep}
\end{figure*}

Figure~\ref{fig-dblangsep} displays the distribution of $\sim 45,000$ isolated pairs of
{\it FIRST} sources as a function of their separation. The minimum value of
$3^{\prime\prime}$ is imposed by the source detection algorithm (WBGH).
The fall-off in the number of detected doubles beyond $\sim
10^{\prime\prime}$ reflects a number of factors: 1) the $5^{\prime\prime}$
survey beam resolves out very extended features and will therefore not
detect sources on the largest scales, 2) the exclusion of triples and other
multiple component sources from this subsample will preferentially remove
extended sources on larger scales, and 3) there is a
real decline in the number of large angular-diameter objects. For comparison
with the APM catalog, we have selected objects with separations in the range
$8^{\prime\prime}<d< 30^{\prime\prime}$.

\begin{figure*}
\epsscale{0.45}
\plotone{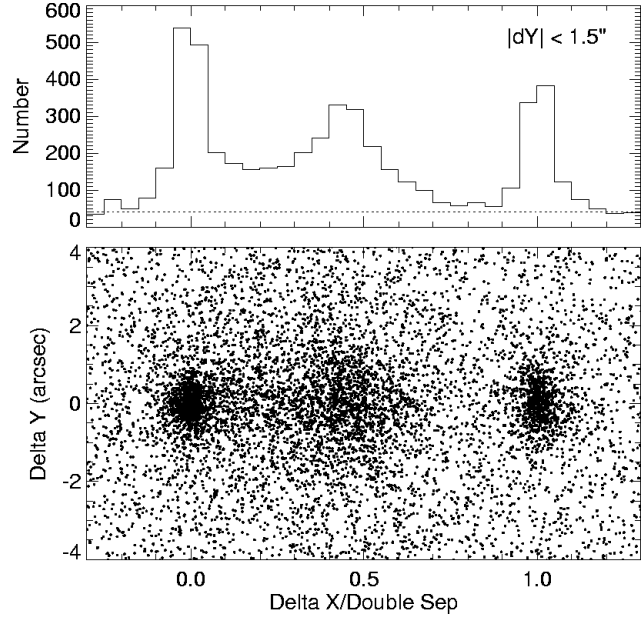}
\epsscale{1}
\caption{
The distribution of the 8220 optical objects
in the vicinity of the 21,579
isolated doubles with separations between $8^{\prime\prime}$ and $30^{\prime\prime}$. The
axes are defined such that the line joining the two component centroids
is the x-axis and the normal to this line is the y-axis. Distances
along the x-axis are normalized to the component separation with the
origin defined to coincide with the brighter of the two components; the
y-axis is in arcseconds. The lower panel shows the distribution of all
optical objects, while the upper panel displays a histogram of the 5348 objects
falling within $|y|<1.5^{\prime\prime}$. The false rate is shown as the horizontal
dashed line.
}
\label{fig-dblpos}
\end{figure*}

The lower portion of Figure~\ref{fig-dblpos} displays the distribution of optical objects
in the vicinity of the 21,579 pairs of radio sources meeting this criterion.
The x-axis for each pair is defined
by the line joining the centroids of the two objects; the scale is normalized to the
separation, with the origin chosen as the brighter of the two components. The
upper panel shows a histogram for all optical objects within $1.5^{\prime\prime}$ of
the line joining to two components; the expected false rate from a uniform distribution of sources with the
same mean surface density as the {\it FIRST} survey is shown as the
dashed line. Several features of the distribution are
immediately apparent. There is a large concentration of optical objects coincident
with the brighter of the two radio components, suggesting a core-jet
morphology. A roughly equal number of identifications is found approximately
half way between the two components with a slight bias in the direction of
the brighter
lobe; the broader spread in the y-direction for these counterparts reflects in
part the bending of radio lobes as a consequence of their interaction with
the intergalactic medium ({\it e.g.}, Blanton et al. 2000, 2001). Finally, a much smaller
fraction of the identifications is coincident with the weaker of the two lobes.
The overall identification rate derived from integrating the upper curve and
subtracting the background rate is $\sim19$\%, similar to
that for the radio sample as a whole.

\begin{figure*}
\plottwo{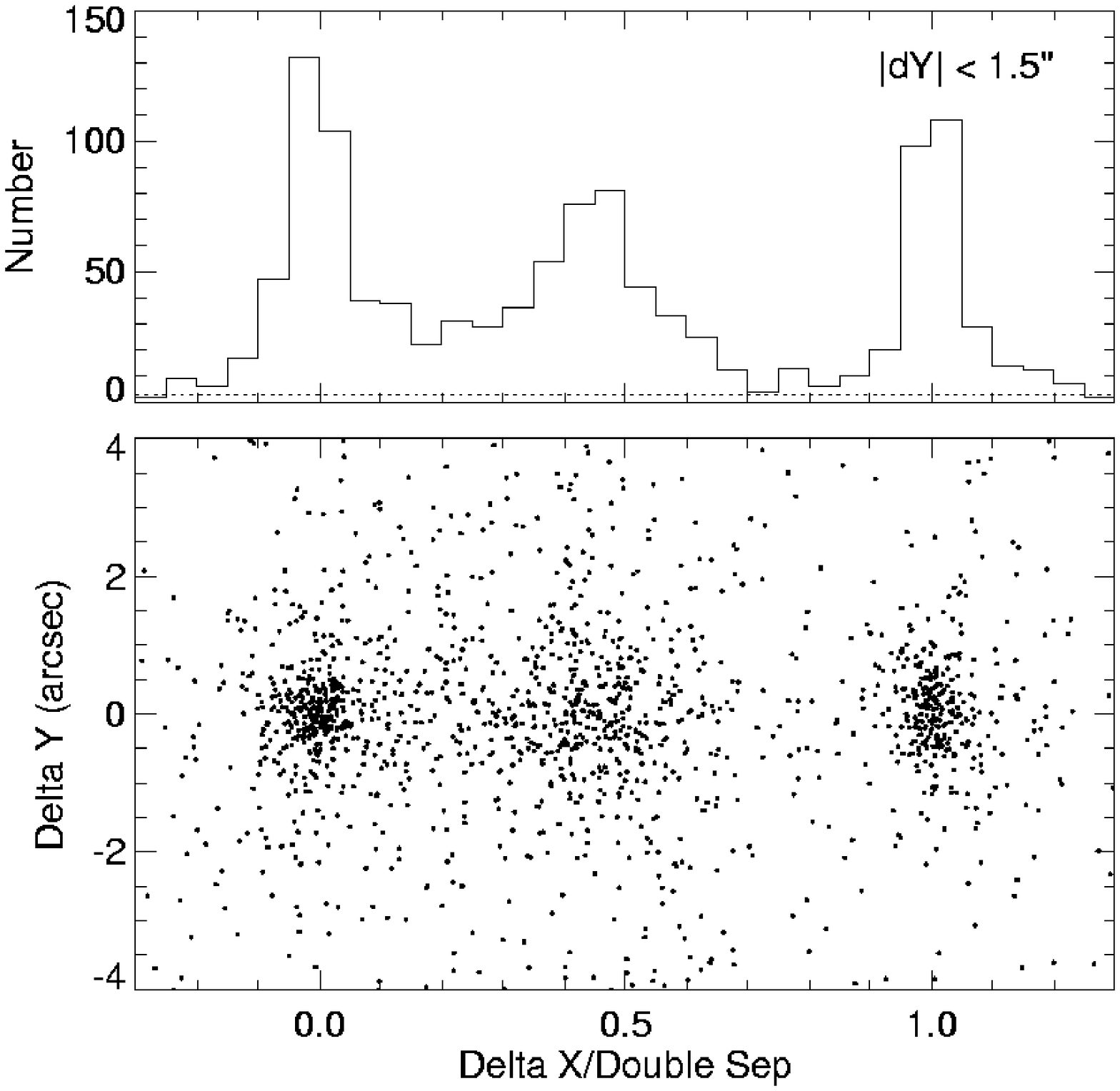}{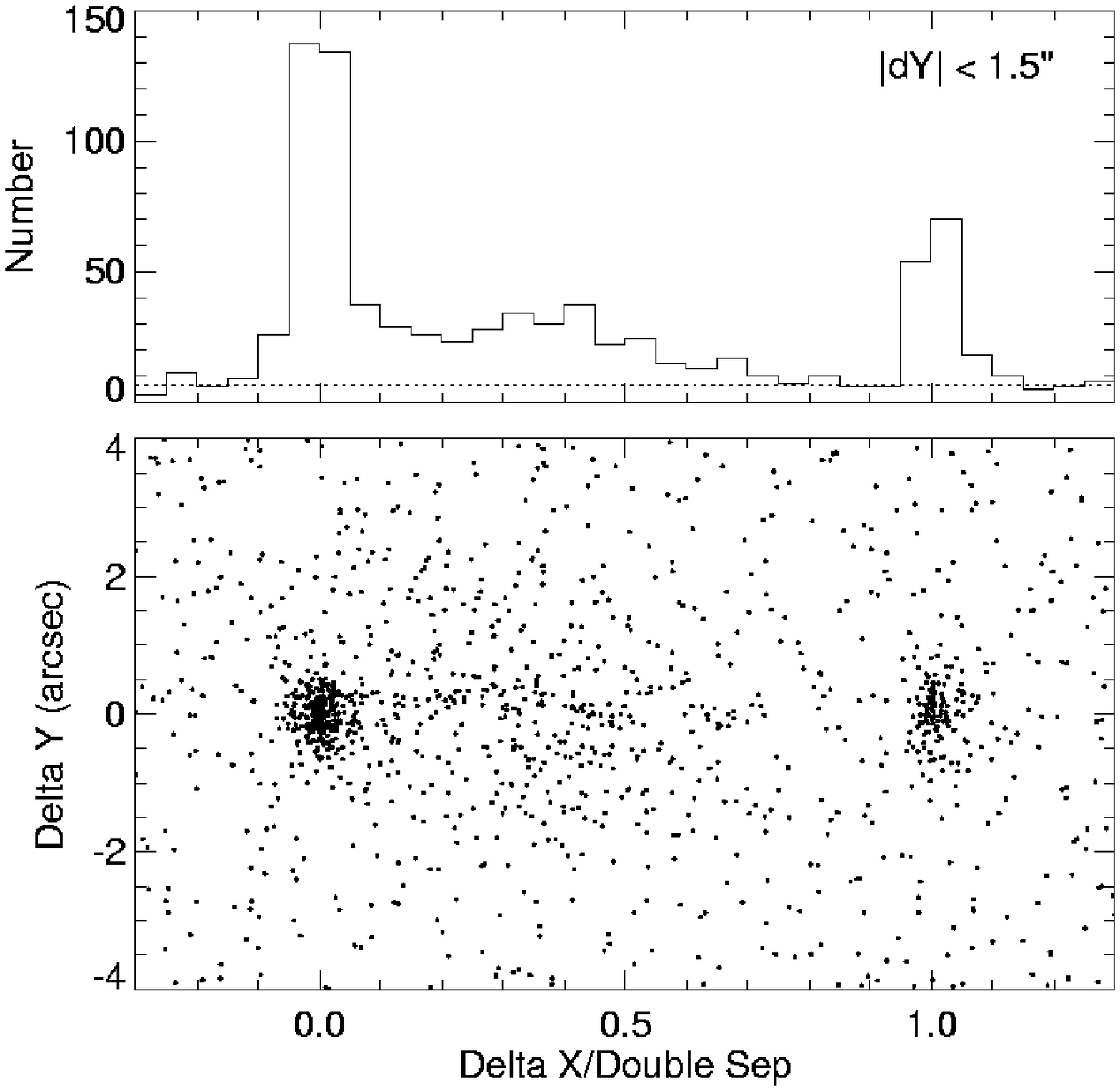}
\caption{
Using the format of Fig.~15, we display the results for
optical counterparts to radio doubles divided between (a) the 1492 objects 
classified as galaxies on both plates and (b) those 1278 objects
classified as stellar on both plates.
}
\label{fig-dblpos_star}
\end{figure*}

\begin{figure*}
\plottwo{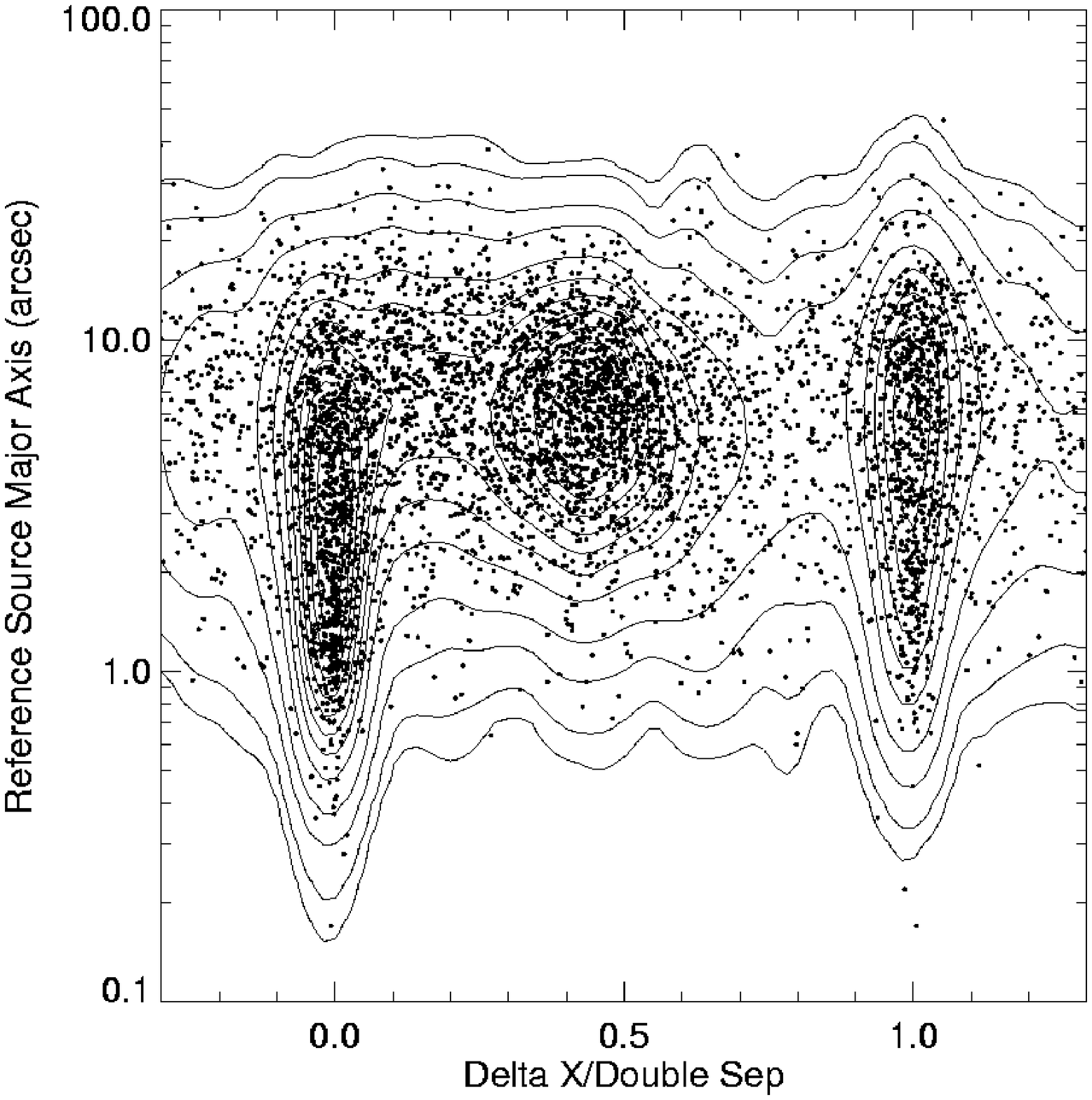}{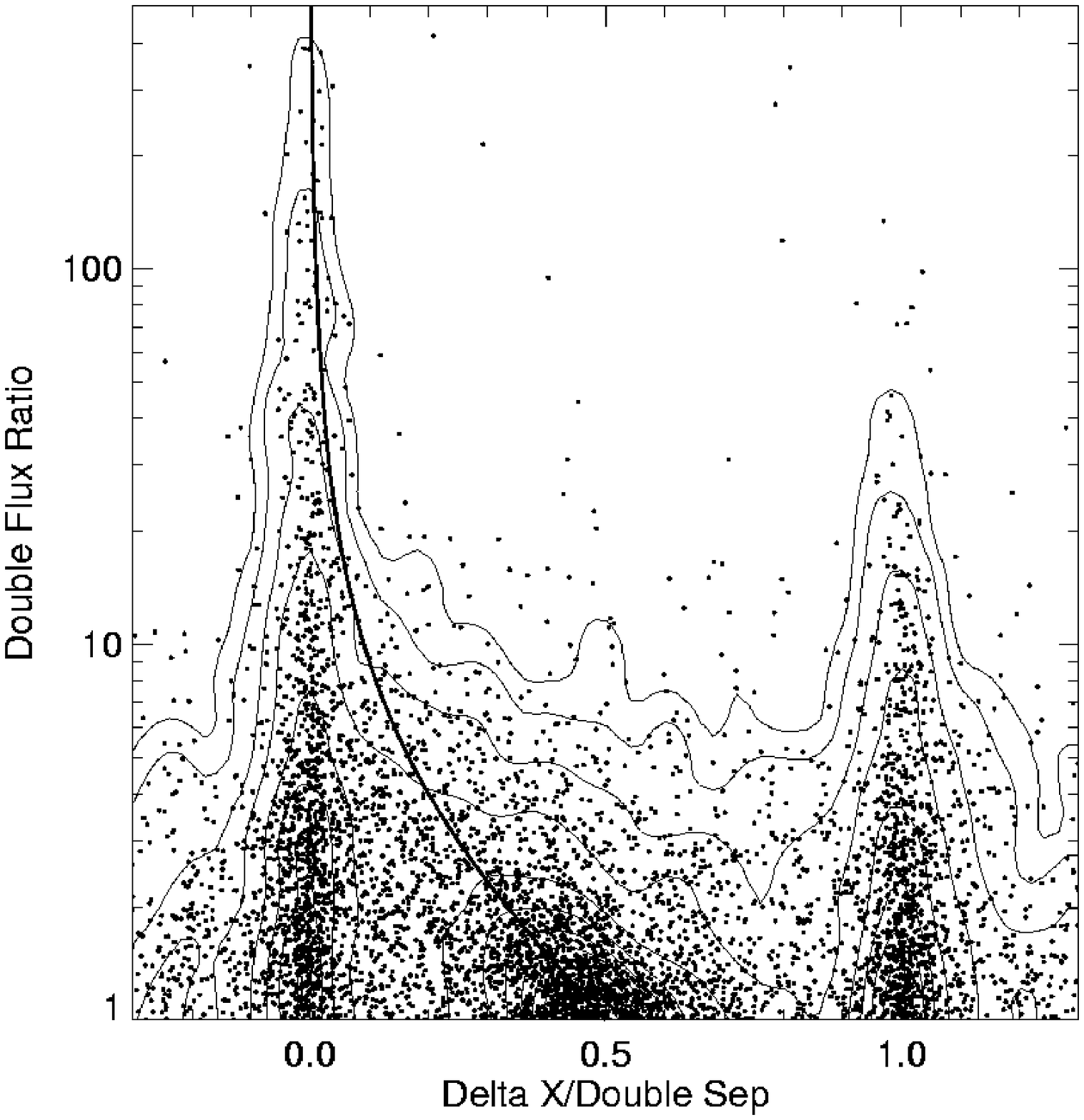}
\caption{
(a) The angular size of the brighter radio component for the 5348
doubles that have optical counterparts within $1.5^{\prime\prime}$ of the line
joining them, versus the distance along that line
(normalized to the double separation as in
Fig.~\ref{fig-dblpos}). Contour lines illustrate the density of points in various
parts of the diagram at levels of 0.01, 0.02, 0.05, 0.1, 0.2, 0.3, \dots, 0.9 of the peak.
A source is clearly resolved in the {\it FIRST} survey if it
has an angular size greater than 2.5--$3^{\prime\prime}$. (A total of 125 objects with
deconvolved sizes of $0.0^{\prime\prime}$ are omitted from the plot; 54 fall near $dx=0$,
with 51 near $dx=1$ and only 12 lying between the components.) The brighter of the
two components (which defines the x-origin) is more often point-like. (b) The
ratio of the flux densities of the radio doubles with counterparts as
a function of distance along the line joining the two components. In cases
where the optical counterparts coincide with one component or the other,
the components tend to have very unequal flux densities, whereas for cases when
the optical counterpart is between the two components, the doubles are of
nearly equal flux density. The thick line indicates the flux-weighted centroid of 
the source.
}
\label{fig-dblfunc}
\end{figure*}

Dividing the optical counterparts by morphological class yields additional
interesting results. The objects classified as galaxies on both plates (Figure
16a) produce the most counterparts midway between the components, while the
stellar objects are predominantly associated with the brighter radio component
(Figure~\ref{fig-dblpos_star}b). Figure~\ref{fig-dblfunc} shows that many of these core components are small in size
(panel a) and are much brighter (panel b) than the other component: the median
flux ratio for sources where the optical counterpart matches the brighter
radio component is 3.0, while the median for the central-component
matches is 1.4. When the brighter of the radio components identifies the optical
counterpart, it also tends to be the smaller of the two components
(panel c). Likewise, for cases in which the dimmer source has the optical
counterpart, it tends to be smaller. In contrast, the sources with identifications
that lie between the two radio lobes have very narrow ranges of flux density
and size ratios near unity, and are virtually all resolved.

\begin{figure*}
\figurenum{17}
\epsscale{0.45}
\plotone{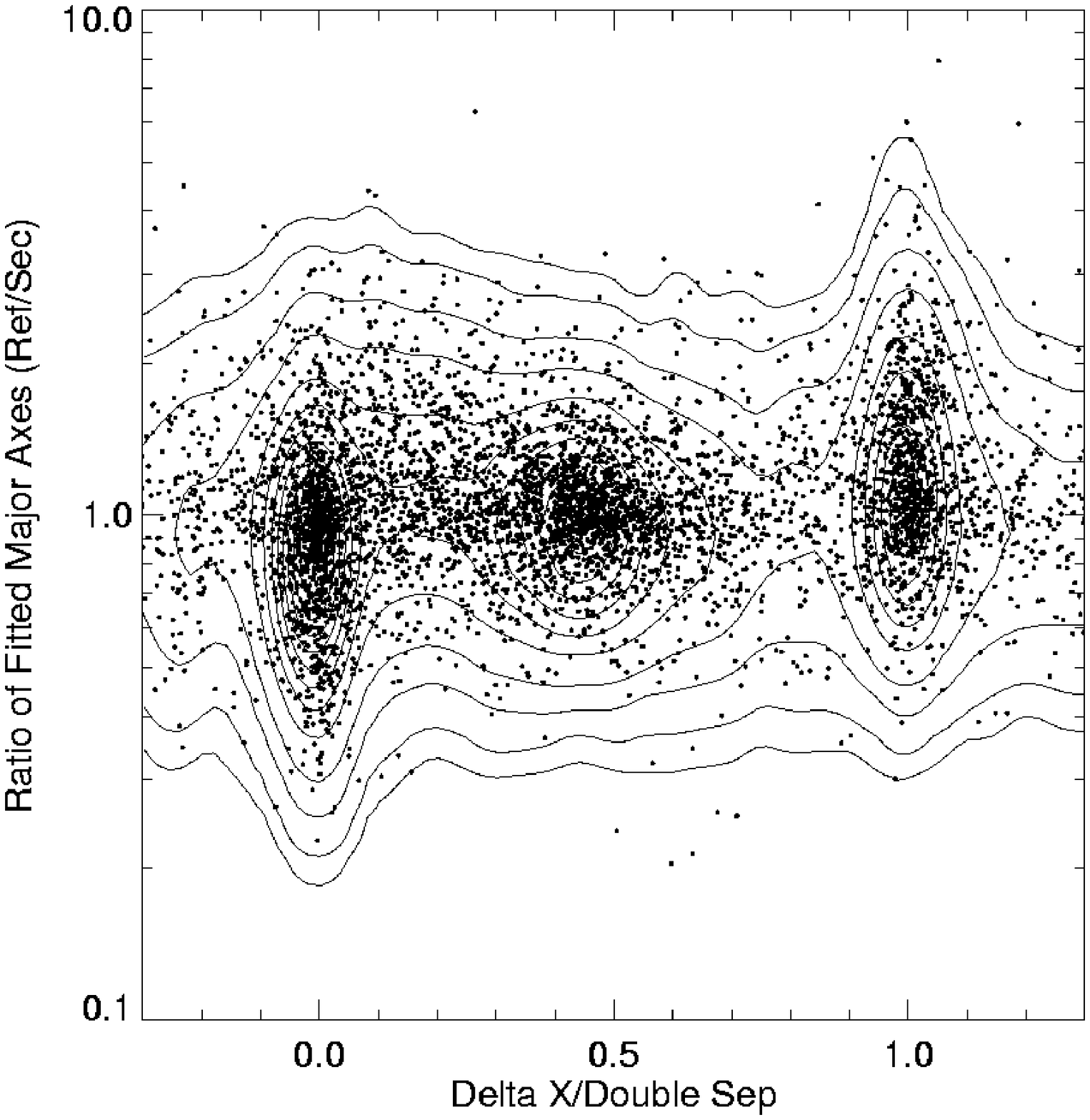}
\epsscale{1}
\caption{
(c) The ratio of the fitted major axes of
the doubles with counterparts as a function of distance along the line joining the
components, defined as the size of the brighter component over the size of the
dimmer one. When the optical counterpart coincides with one component or the
other, the size of that component is smaller, whereas when the
counterpart is found between the two components, they tend to be of
equal size.
}
\label{fig-dblfunc-c}
\end{figure*}

Using these trends in radio source component flux density
and size ratios, total extents,
and optical morphology (in addition to magnitude and color, perhaps) it would
be possible to develop a reasonably reliable identification algorithm for double
and multiple {\it FIRST} sources, although we regard such an effort as beyond the
scope of this paper. 
What is clear from the foregoing analysis is that, since
the identification rate for complex sources is similar to that for
single-component objects, the effect on the final fraction of radio emitters
identifiable at the POSS-I plate limit simply scales with the number of radio
components removed from consideration as a consequence of their association
with another catalog entry. Using an algorithm that assigns a probability to
component associations based on flux density and separation, we find that,
for separations up to $120^{\prime\prime}$ there are 32,312 doubles\footnote
{Note this number is different from that in Figure 15 because the latter
includes only {\it isolated} doubles so as to examine as cleanly
as possible the optical coincidence rates.}, 12,802 triples, and 5,716
groups of four or more sources with a probability of real association $>90\%$.
This reduces our catalog to $\sim 306,000$ discrete radio emitters and yields
an overall identified fraction at the POSS-I limit of $\sim24$\%.

\section{Characteristics of the Optical Counterparts to FIRST Sources}

The objects in the {\it FIRST} catalog span over four orders of magnitude
in radio flux density, and the identification rate as a function of flux
density is displayed in Figure 18. Even for sources as bright as 1~Jy, only
one-third of the objects have counterparts brighter than the plate limit.
The identification rate falls monotonically with flux density to 20~mJy
where it reaches a minimum of $\sim 12$\% and then climbs again toward the
survey limit. The rise at low flux densities reflects the change in the
composition of the radio source population which is dominated by distant
AGN at high flux densities and by nearby star-forming galaxies below
1~mJy (Condon 1992). The fraction of sources classified as stellar on at
least one of the two POSS plates is also reflective of this trend (Figure 19).
Above 100~mJy over 80\% of all counterparts are stellar (quasars),
whereas this fraction decreases
monotonically until at 1~mJy, 60\% of the counterparts are classified as
galaxies; given the strong tendency toward the misclassification of faint
galaxies as stellar, the true fraction of galaxy counterparts is
considerably higher.

\begin{figure*}
\epsscale{0.45}
\plotone{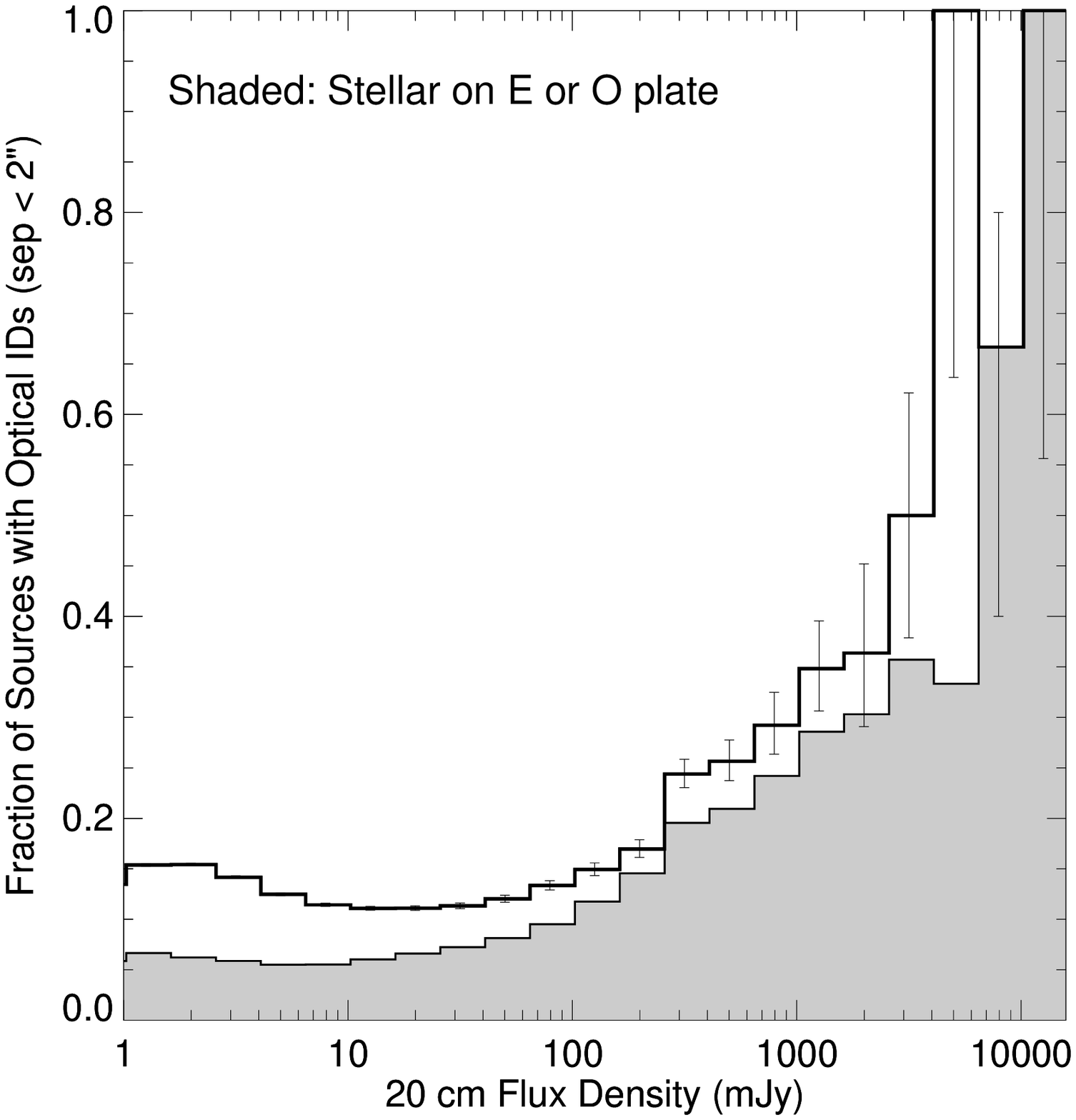}
\epsscale{1}
\caption{
The fraction of {\it FIRST} sources with optical counterparts (matching to
within $2^{\prime\prime}$) as a function of radio flux density.  The shaded
histogram includes objects classified as stars on at least one of
the POSS-I plates; the remaining objects are classified as
galaxies.
}
\label{fig-optfrac}
\end{figure*}

\begin{figure*}
\epsscale{0.45}
\plotone{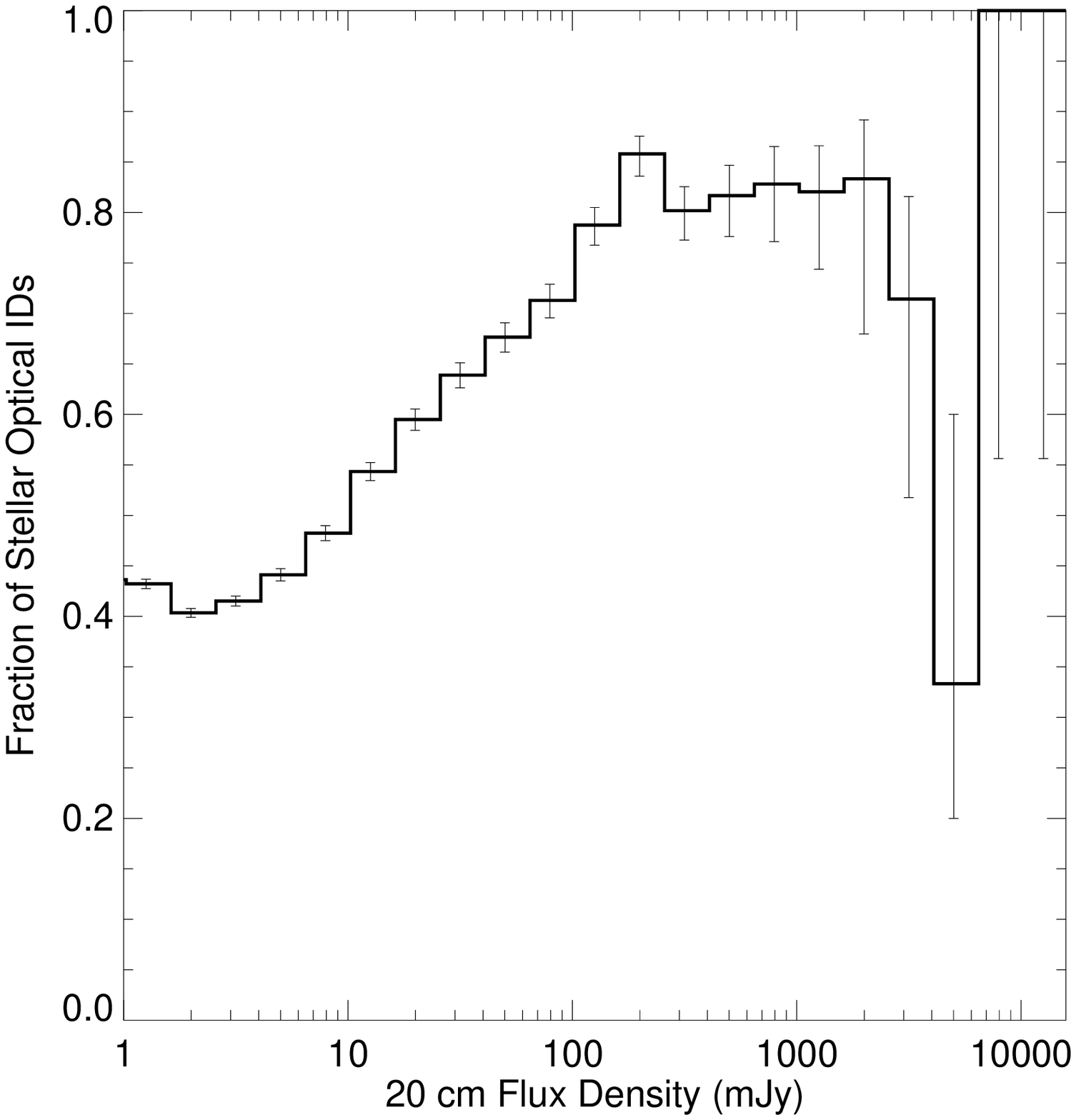}
\epsscale{1}
\caption{
The fraction of {\it FIRST} sources with stellar optical counterparts (matching
to within $2^{\prime\prime}$) as a function of radio flux density. Given the tendency
toward misclassification of faint galaxies as stellar, the true fraction
of galaxy counterparts at faint flux levels is considerably higher than 60\%.}
\label{fig-optstarfrac}
\end{figure*}

One way to explore the types of objects associated with
{\it FIRST} sources is to place the optical counterparts onto a color -
magnitude diagram.  In Figures 20 and 21 we show the color-magnitude
diagrams for the `stars' and galaxies (based on a consistent APM
classification from both POSS plates) within $1.5\arcsec$ and $4\arcsec$
match radii, respectively.
For comparison, these figures also display similar diagrams for an equal
number of stellar and nonstellar sources taken from random
positions on the same plates. 

\begin{figure*}
\plottwo{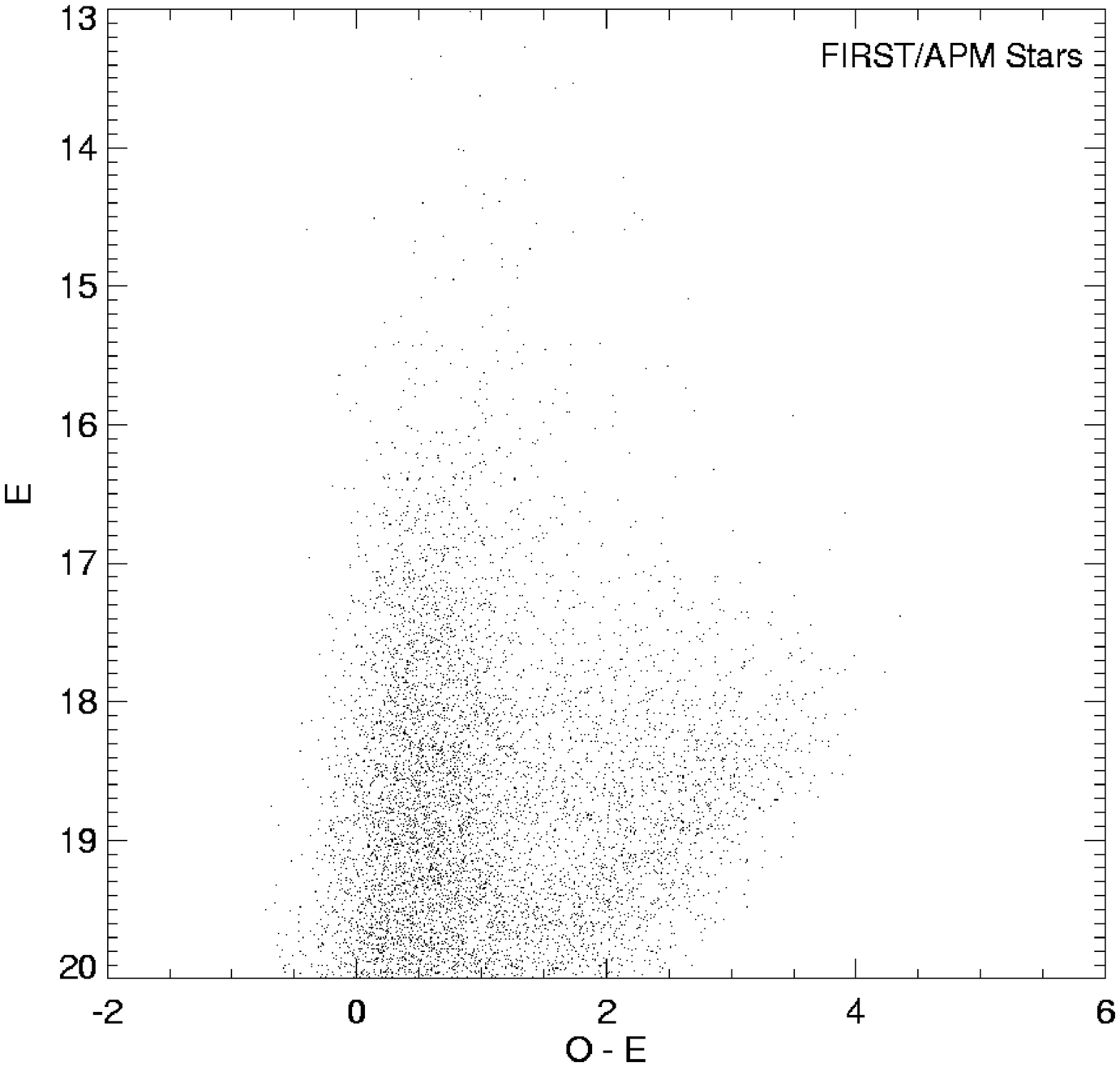}{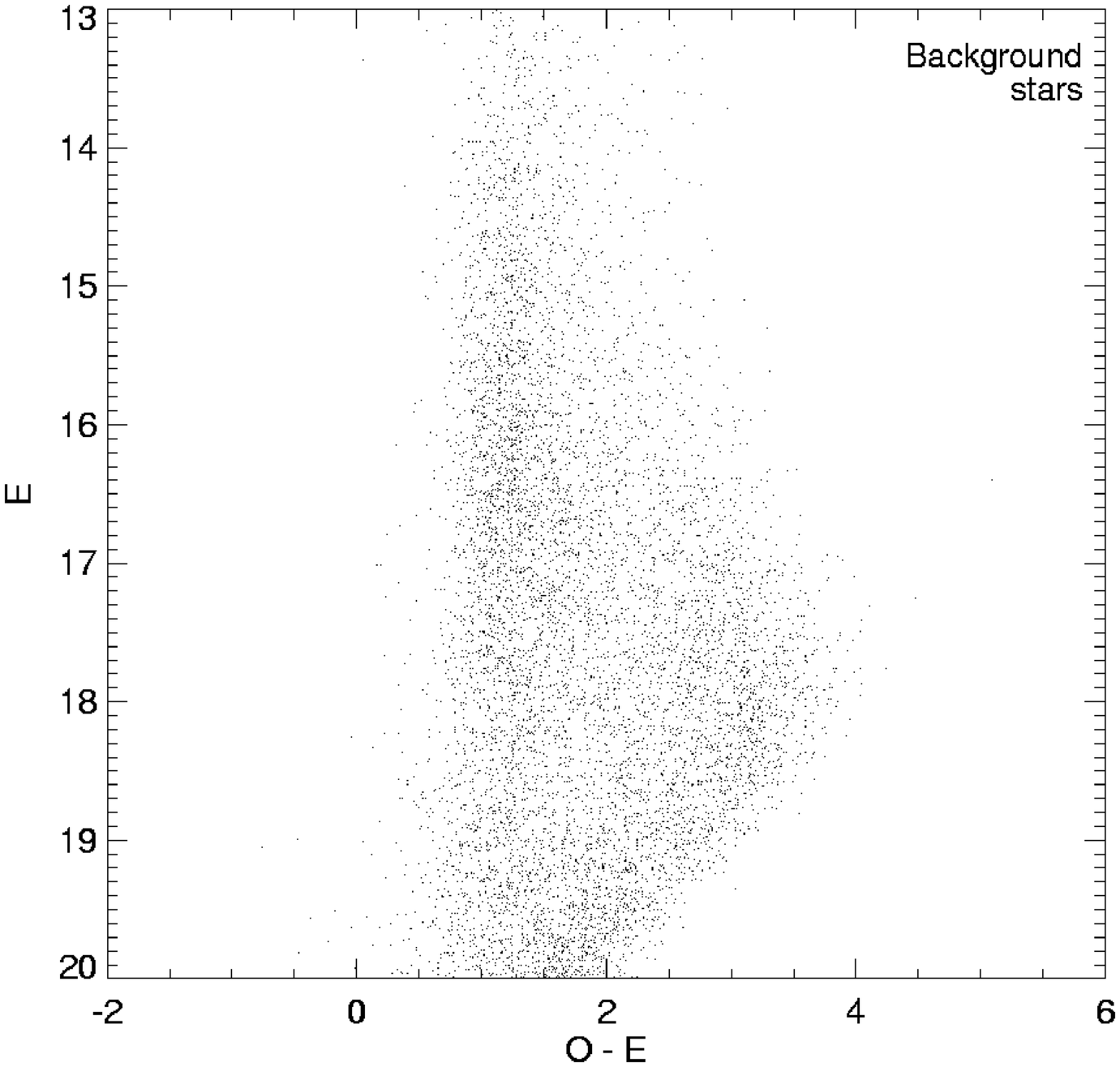}
\caption{Color-magnitude diagrams for the 7877 objects classified as stellar on
both plates in the APM catalog. (a) Objects lying within $1.5^{\prime\prime}$ of
{\it FIRST} sources. (b) An equal number of stellar objects randomly
selected from the same plates. The {\it FIRST} counterparts, fainter and
bluer than the typical star, are mostly quasars.
}
\label{fig-cmplot7s_pub.ps}
\end{figure*}

\begin{figure*}
\plottwo{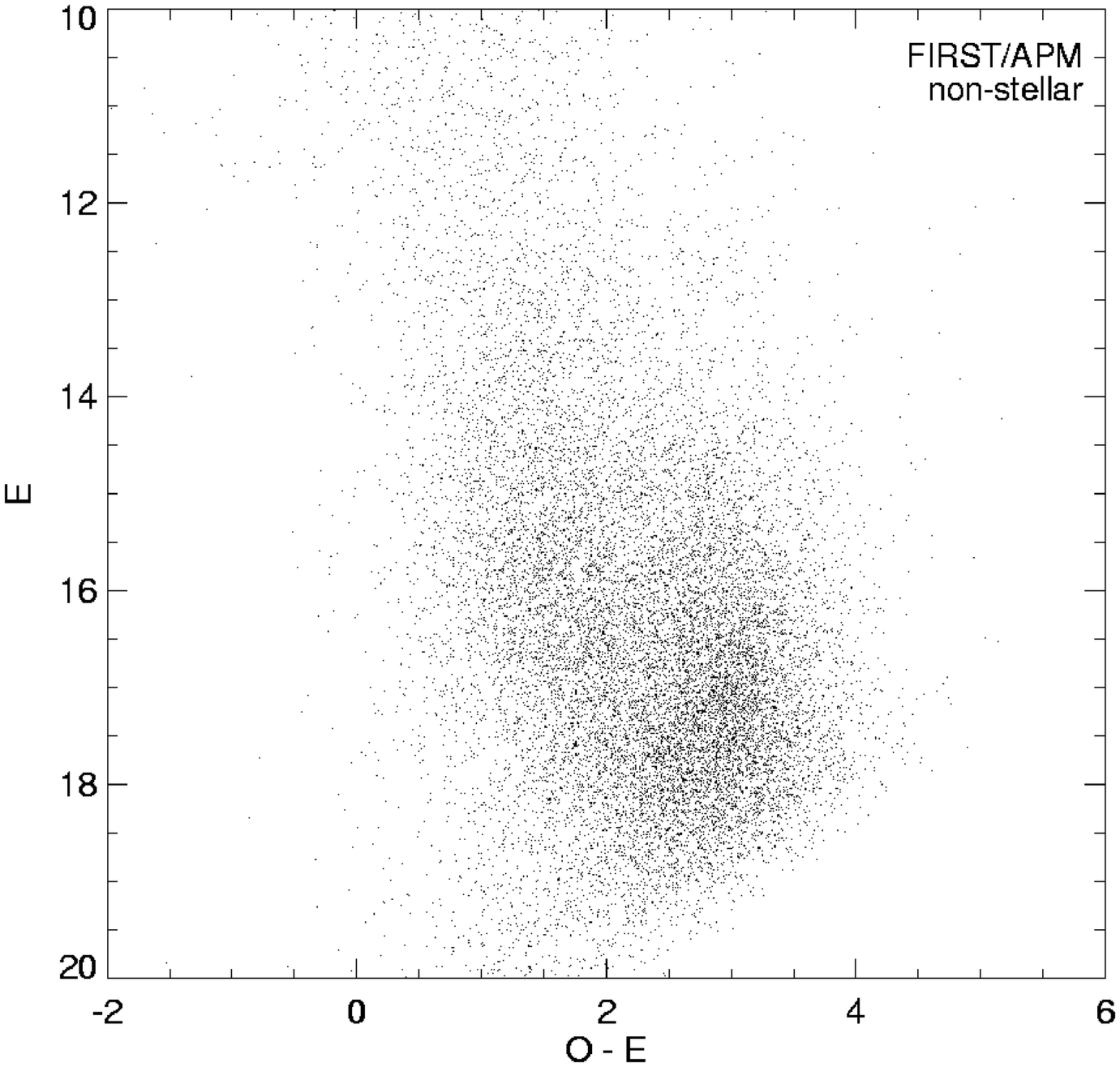}{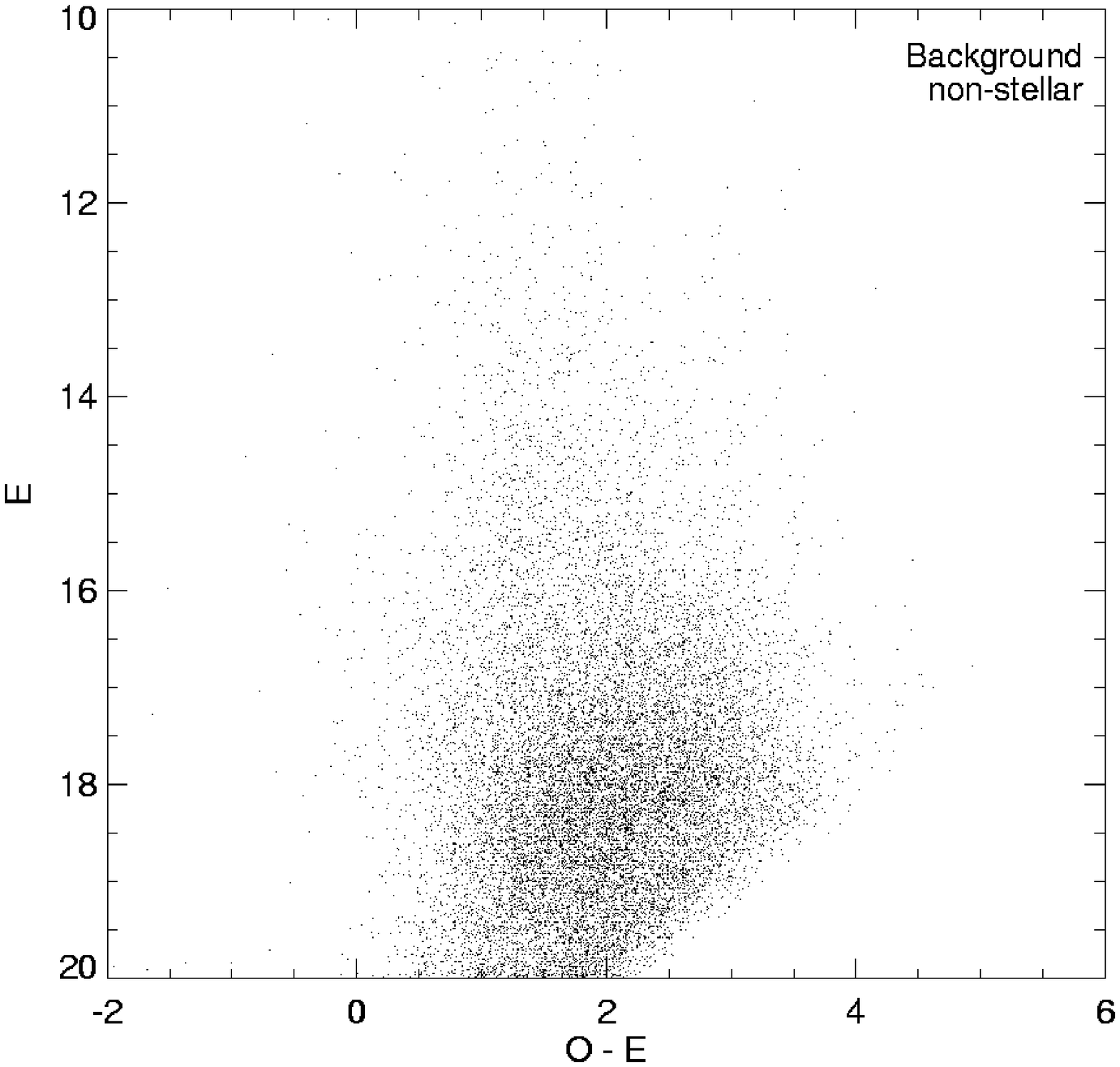}
\caption{Color-magnitude diagrams for the 16,047 objects classified as nonstellar on
both plates in the APM catalog. (a) Objects lying within $4^{\prime\prime}$ of
{\it FIRST} sources. (b) An equal number of nonstellar objects randomly
selected from the same plates. The {\it FIRST} counterparts are, on average,
brighter.
}
\label{fig-cmplot7g_pub.ps}
\end{figure*}

The radio counterparts classified as stellar
are clearly much fainter and much bluer than most `stars'; this is
illustrated in Figure 22, where we plot the fraction of the
radio-detected `stars'
as a function of location in the color-magnitude plane. The detected
fraction peaks at $\sim10$\% for colors around $O-E=0$ for $18<E<20$. These
objects are nearly all quasars. The increasing fraction of detected objects
with all colors fainter than $18^{th}$ magnitude is largely an artifact
produced by the failure of the classifier to distinguish stars from
galaxies near the plate limit; in fact, most of the faint, red stellar
objects are actually galaxies, although a handful of very red objects
have been shown to be high redshift quasars (Hook et al. 1998).

\begin{figure*}
\epsscale{0.45}
\plotone{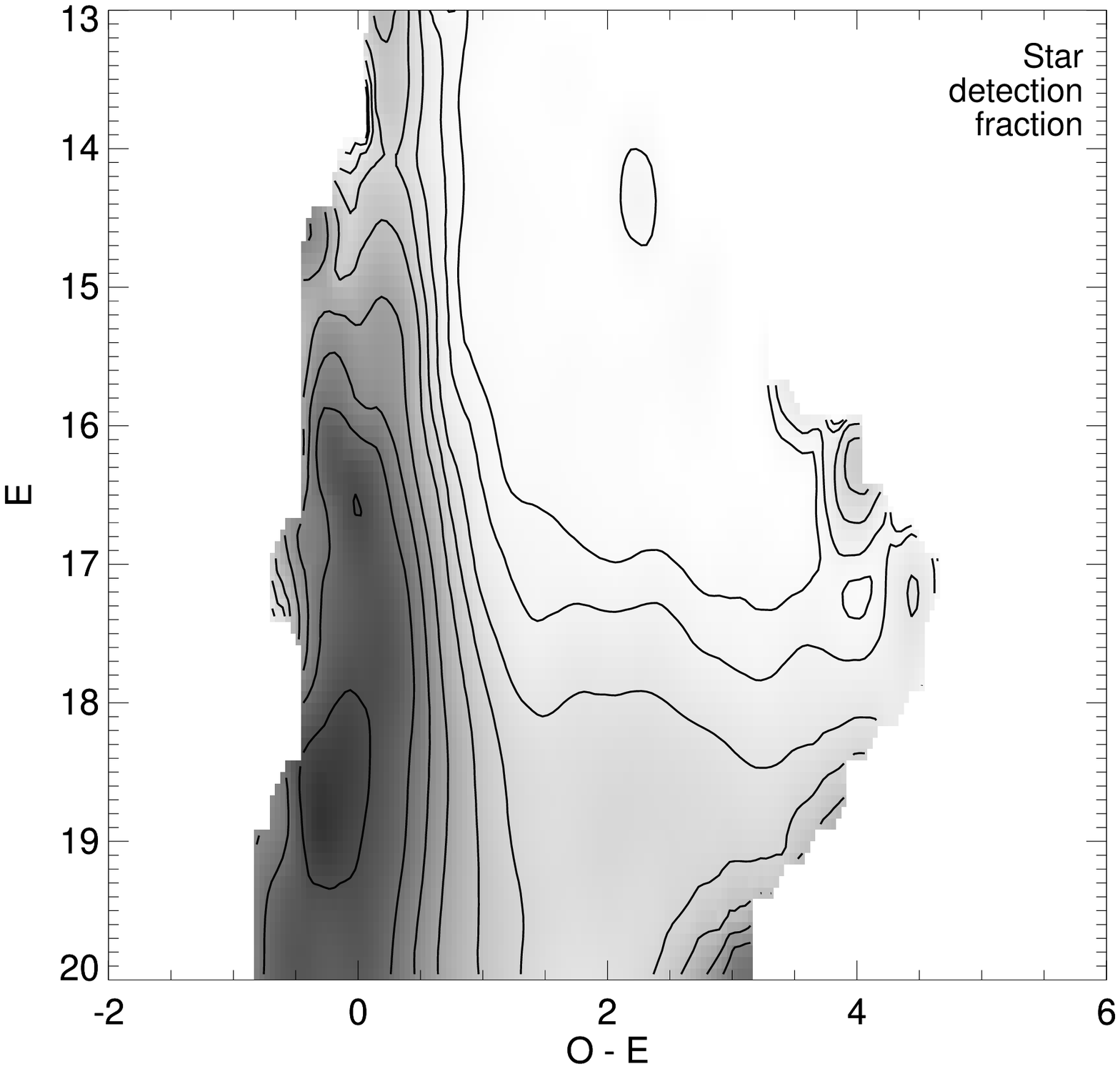}
\epsscale{1}
\caption{A contour and greyscale display of the fraction of objects
classified as stellar on both plates which have radio counterparts.
Darker regions have higher detected fractions.
The fraction reaches nearly 10\% for colors of $O-E=0$. Contour
levels are 0.0001, 0.0002, 0.005, 0.001, 0.002, 0.005, 0.01, 0.02, and 0.05.
}
\label{fig-cmplot4s_pub.ps}
\end{figure*}

For the nonstellar objects, it is clear that the radio detections are
biased toward brighter magnitudes, but are found for galaxies of
all colors. In Figure 23, we show the detected fraction
of galaxies in the color-magnitude plane. The roughly uniform detection rate as
a function of color and the bias towards brighter galaxies are apparent; up
to 10\% of the brightest galaxies are detected and even for E$>$16, $>$2\% of
all galaxies have {\it FIRST} counterparts.

\begin{figure*}
\epsscale{0.45}
\plotone{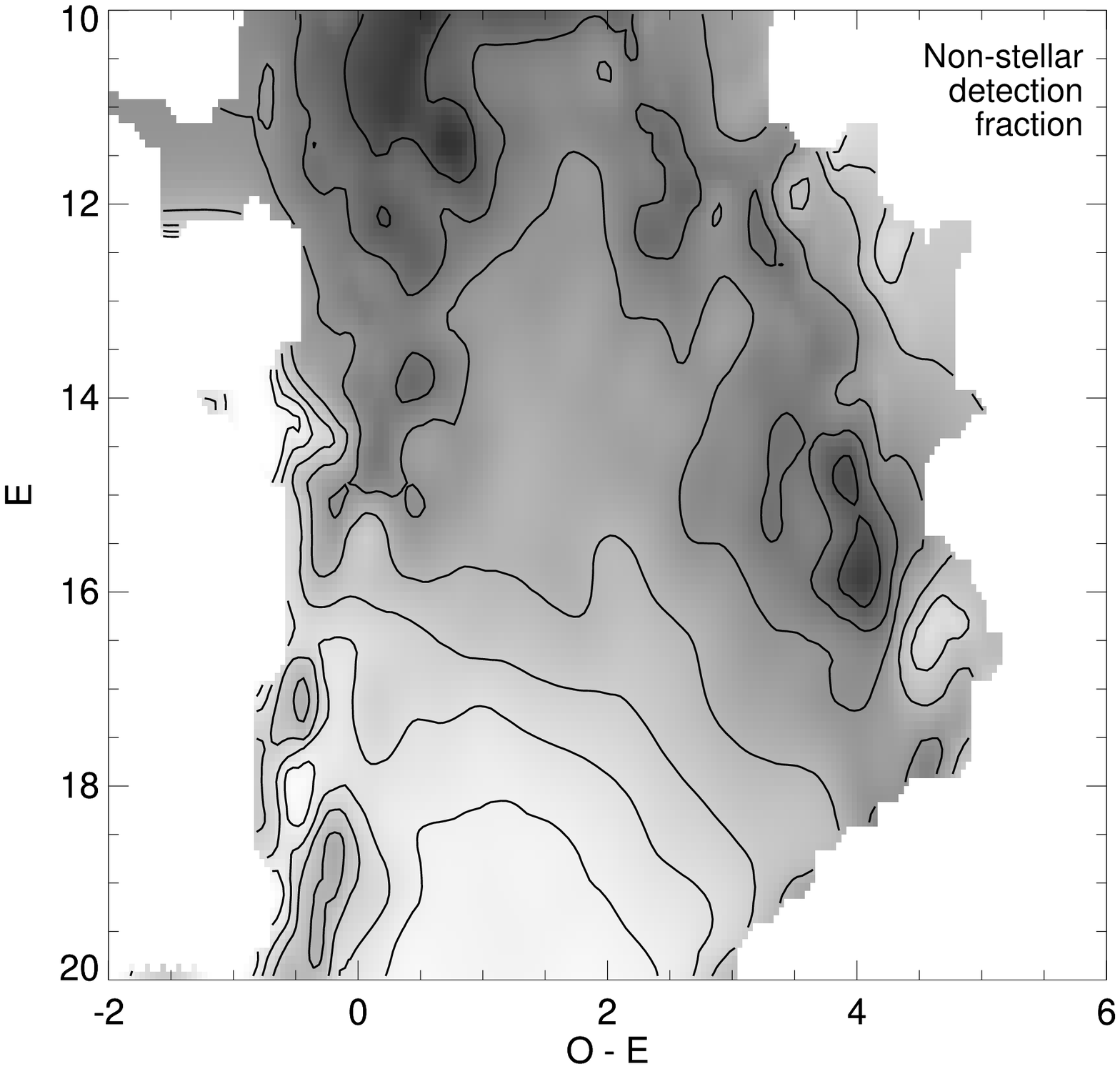}
\epsscale{1}
\caption{A contour and greyscale plot of the fraction of objects
classified as nonstellar on both plates which have radio counterparts.
More than 10\% of all galaxies are detected at the brighter magnitudes.
Contour levels are 0.002, 0.005, 0.01, 0.02, 0.05, 0.1, and 0.2.
}
\label{fig-cmplot4g_pub.ps}
\end{figure*}

\section{The Catalog of FIRST Optical Counterparts}
\label{sectioncatalog}

Table~4 displays a sample page of the catalog of APM-{\it FIRST}
matches which is available from the {\it FIRST} home page
(\url{http://sundog.stsci.edu}).  Columns 1 and 2 contain the radio
source coordinates (epoch J2000.0) and are followed by the offsets in
Right Ascension, Declination, and radius (all in arcsec) of the nearest
optical object falling within $5\arcsec$ of the radio position. The
global APM/{\it FIRST} shift, the plate-to-plate shifts, and the
intraplate shifts as determined by {\it FIRST} (\S4)
have all been applied to the raw APM catalog positions before
calculating these offsets.
Columns 6--8 contain the optical classification, the psf parameter
(see \S3), and the magnitude of the counterpart on the red ($E$)
plate. The optical classification codes are: stellar (consistent with
the magnitude- and position-dependent point spread function, cl$=-1$),
non-stellar (a measurably extended source, cl=1), merged objects
(sources with two local maxima within a single set of connected
above-threshold pixels, cl=2), and noise (objects with nonphysical
morphologies, cl=0). For further details of the principles involved,
see \S3 and Maddox et al.  (1991a,b). A negative magnitude indicates
that the object was not detected in this waveband and the absolute
value is a 2$\sigma$ upper limit.  Columns 9--11 list the same
information for the blue ($O$) plate.  Column 12 is the ($O-E$) color
derived from columns 8 and 11 unless the object is only detected on
one plate, whereupon the color is designated -9.99.  All raw APM
magnitudes have been corrected for plate to plate systematic
calibration errors as described in \S 3.4; the uncertainties (rms) in
the magnitudes are $\sim0.2$.  The POSS plate number in column 13 is
followed by selected radio parameters for the {\it FIRST} source,
including the peak and integrated flux densities, and the deconvolved
source major axis, minor axis, and position angle.  The major and
minor axes (FWHM) have been deconvolved using a 5.4'' synthesized beam.
Negative values indicate that the source size (before deconvolution) was
smaller than the beam.
These radio parameters are described in further detail in WBHG.

As demonstrated above, 95\% of the counterparts within $2\arcsec$ are
actual {\it FIRST} identifications. There are, however, a number of
reasons for studying the contents of the optical sky even farther from
the {\it FIRST} positions: double and multiple-component sources are
likely to have parent galaxies with substantial offsets from
the radio positions, and the clustering of optical objects around {\it
FIRST} sources appears to be a fruitful area for investigation (Helfand {\it et
al.} 1998). In addition, catalog users may find it helpful to be
able to calculate false rates for particular ranges of optical
magnitude, radio flux density, or source morphology. Thus, we have
constructed a second catalog (also available on-line) which includes
{\it all} optical objects appearing on either POSS-I plate which lie within
$20\arcsec$ of a {\it FIRST} position;  68\%  of the {\it FIRST}
sources have at least one such counterpart. The format of this database
is identical to that of the primary catalog; it contains nearly half a million
entries representing 454,754 unique optical objects in the vicinity of
261,434 radio catalog entries.
Also as the FIRST survey grows in size we shall periodically update these 
on-line optical catalogues.

\section{Discussion}
\label{sectiondiscussion}

The systematic optical identification of radio-selected samples with
flux densities in the range 1--30~mJy began nearly twenty years
ago using the first deep radio images obtained by Westerbork and the
VLA.  In Table~\ref{table-ids}, we summarize the published work on this problem. The
area surveyed covers a little more than 50 deg$^{2}$ and includes 1448
radio sources between the limits of the all-sky, single dish surveys
($\sim$30 mJy) and the {\it FIRST} survey threshold (1.0 mJy). A total
of 455 sources have suggested identifications, albeit to different
levels of optical sensitivity, completeness, and reliability. Only 270
counterpart candidates have been published for radio-selected objects
at thresholds roughly equivalent to the POSS-I plates. The total number from
this flux density range in
the work reported here represents a sample of $\sim65,400$ confidently identified radio
sources to this limit. Our project, then, represents an increase of more
than two orders of
magnitude in the number of optically identified faint radio sources.
What source classes are represented in this sample, and how do the
results compare with our expectations from earlier work?

The raw fraction of detections at the POSS-I limit, $\sim 18.3\%$, is similar
to the mean value derived from previous work (Table~\ref{table-ids});
the different optical bands used and the fuzzy lower bound to the
POSS-I magnitudes easily explains the small discrepancy introduced when
the corrections for multiple-component sources  are applied (\S 5.4). The
distribution amongst stellar and non-stellar counterparts is also roughly
similar; detailed comparisons are not warranted given the inevitable
difficulty of classifying objects near the plate limits where most of the
candidates lie. Further progress in quantifying the source populations
represented, and in establishing their luminosity functions, requires CCD
imaging and spectroscopy, work we have begun to pursue for a variety
of subsamples derived from this {\it FIRST}/APM database. We discuss
these briefly here.

The comprehensive catalog of stellar radio observations by Wendker
(1995) contains 18 stars in the {\it FIRST} survey area detected at any
radio frequency between 100 MHz and 30 GHz;  of these, only a dozen
have ever been seen at a 20~cm flux density in excess of 1 mJy.  Our
comprehensive study of various stellar proper motion catalogs
has produced an additional fourteen stellar radio source
identifications (Helfand et al. 1999). Proper motions remain the
largest obstacle to our identification of stellar counterparts fainter
than Tycho catalog limit ($\sim11^{th}$ magnitude), and the advent of
the GSC-II should significantly increase the number of stellar identifications
to $\sim85$. Nonetheless, stars make a trivial contribution to the
total radio source population above our 20~cm flux density threshold of
1 mJy. Other Galactic counterparts are similarly rare: we have detected
only 3 (of the 8) known planetary nebulae and 7 (of the 18) known
radio pulsars. In total, we expect that less than 0.03\% of the
{\it FIRST} catalog entries represent Galactic objects.

There are, of course, a much larger number of optical counterparts with
stellar images on the POSS plates. These represent a mixture of
quasars, BL Lac objects, AGN with sufficiently bright nuclei that they
are classified as stellar,  H~II galaxies with similarly bright nuclei,
and radio galaxies with sufficiently small diameters that they are
unresolved. The {\it FIRST} Bright Quasar Survey (White et al. 2000)
has established the distribution amongst these categories for objects
brighter than E=17.8: $\sim55\%$ quasars, $\sim8\%$ BL Lacs, 
$\sim9\%$ AGN, $\sim 4\%$ radio galaxies, and $\sim17\%$ H~II galaxies,
with the remaining objects being a collection of stars and chance coincidences.
We have begun a smaller survey to establish this distribution at fainter
magnitudes (Becker et al. 2001).

The majority of the {\it FIRST} counterparts brighter than the POSS-I limit
are galaxies. Two principal populations are represented: active
galactic nuclei including both relatively nearby Seyferts and luminous
elliptical hosts of more powerful radio galaxies up to $z\sim0.5$, and
a local population of star-forming galaxies (Condon 1992). Our work to
date has consisted primarily of following up the counterparts to the several
hundred bent, double-lobed radio galaxies which can act as tracers for 
galaxy clusters (Blanton et al. 2000, 2001).
Studies of selected, magnitude-limited samples of the galaxy population
are in progress.

One important use of {\it FIRST} radio sources is as tracers
of the general matter density of the Universe; their broad redshift
distribution and high mean $z$ samples a much larger volume than most
surveys in other wavelength regimes. In applications such as determining the
two-point (Cress et al. 1996), and higher order (Magliochetti et al. 1998)
angular correlation functions, searching for a weak lensing signal on
large angular scales (Refregier et al. 1999) and other such statistical
studies, it may be useful to eliminate the low-redshift, star-forming galaxy
population. The catalog presented here can be applied for such purposes.

\section{Summary and Future Work}
\label{sectionsummary}

We have completed an optical identification program for {\it FIRST} radio
sources by matching the {\it FIRST} catalog to a catalog constructed from
the APM scans of the POSS-I survey. A detailed analysis of the properties
of the APM catalog has been undertaken; we have recalibrated its photometry
using the APS, and have produced a factor of $\sim10$ improvement in
astrometric accuracy by using the {\it FIRST} sources as a set of astrometric
standards. We have discussed in some detail the procedures for identifying
radio source counterparts through positional coincidence, calculating false
match rates for various subsets of both catalogs. For stellar counterparts
of isolated radio sources, the matches are 90\% reliable and 90\%
complete within an offset of $1.2\arcsec$ ; over 98\% of all matches within
$1\arcsec$ are correct identifications. In total, nearly 73,000 radio
source identifications are recorded, an increase by a factor of
$\sim200$ over the total number of counterparts previously
reported for radio-selected samples in the flux density range 1-30~mJy.
We present color-magnitude diagrams for the radio counterparts and
briefly discuss the various source populations represented.

A catalog of all 454,754 optical objects within $20\arcsec$ of the
382,892 radio sources in the north Galactic cap region {\it FIRST} has
covered to date is available on the {\it FIRST} Webpage. As we complete
radio observations and optical matching for additional regions of the
sky, this catalog will be expanded accordingly. The results presented
here suggest that future work to match {\it FIRST} with the GSC-II
and the Sloan Digital Sky Survey (with which the {\it FIRST} sky coverage
is matched) will yield enormous, valuable samples of optically identified
radio sources for use in a wide variety of applications from
stellar radio astronomy to cosmology.

\acknowledgments

The success of the {\it FIRST} survey is in large measure due to the
generous support of a number of organizations. In particular, we
acknowledge support from the NRAO, the NSF (grants AST-98-02791 and
AST-98-02732), the Institute of Geophysics and Planetary Physics
(operated under the auspices of the U.~S.\ Department of Energy by
Lawrence Livermore National Laboratory under contract No.
W-7405-Eng-48), the Space Telescope Science Institute, NATO, the
National Geographic Society (grant NGS No.~5393-094), Columbia
University, and Sun Microsystems. We thank Juan Cabanela for his
assistance in matching with the APS catalog. RGM acknowledges support
from the Royal Society. DJH is grateful for the
support of the Raymond and Beverly Sackler Fund, and joins RHB and RLW
in thanking the Institute of Astronomy of the University of Cambridge
for hospitality during much of this work.  
We thanks Mike Irwin for his explanatory comments on a draft of this 
paper. This paper is Contribution
Number 685 of the Columbia Astrophysics Laboratory.

{}

\clearpage

\begin{deluxetable}{lcc}
\tablewidth{0pt}
\tablecaption{APM Background Source Density\label{table-background}}
\tablehead{
\colhead{APM Source} & \colhead{$\bar{\rho}_{opt}$}        & \colhead{$\sigma_{opt}^2$} \\
\colhead{Properties} & \colhead{($10^{-4}$ arcsec$^{-2}$)} & \colhead{($10^{-8}$ arcsec$^{-4}$)}
}
\startdata
All matches                     & 7.58 &  7.10 \\
Stellar on both plates          & 1.60 &  0.06 \\
Stellar on at least 1 plate     & 5.16 &  3.85 \\
Non-stellar on both plates      & 0.41 &  0.13 \\
Non-stellar on at least 1 plate & 2.98 &  3.57 \\
Detected on both plates         & 2.56 &  0.51 \\
Detected on only 1 plate        & 5.02 &  6.59 \\
Detected only on red plate      & 2.60 &  2.76 \\
Detected only on blue plate     & 2.43 &  3.83 \\
\enddata
\end{deluxetable}

\begin{deluxetable}{llcccc}
\tablewidth{0pt}
\tablecaption{Confidence Radii and False Rates for Various Source Types\label{table-confidence}}
\tablehead{
\colhead{APM Sources} & \colhead{{\it FIRST} Sources} &
\colhead{90\% Radius\tablenotemark{a}} & \colhead{Number of} &
\colhead{Number of Chance} & \colhead{Percentage of}\\
& &
\colhead{(arcsec)} & \colhead{True Matches\tablenotemark{b}} &
\colhead{Coincidences\tablenotemark{c}} & \colhead{False Matches}
}
\startdata
All                          & All        & 4.2 &     65400 &           13000 &     16.6\% \\
Non-stellar\tablenotemark{d} & All        & 4.0 &     22600 &     \phn\phn640 &  \phn2.7\% \\
Stellar\tablenotemark{d}     & All        & 1.7 &  \phn7600 &     \phn\phn520 &  \phn6.3\% \\
All                          & Unresolved & 2.0 &     17800 &     \phn\phn850 &  \phn4.6\% \\
Non-stellar\tablenotemark{d} & Unresolved & 2.2 &  \phn6200 &  \phn\phn\phn60 &  \phn0.9\% \\
Stellar\tablenotemark{d}     & Unresolved & 1.1 &  \phn2300 &  \phn\phn\phn60 &  \phn2.4\% \\
\tablenotetext{a}{APM-{\it FIRST} separation that encloses 90\% of the true associations.}
\tablenotetext{b}{Estimated number of true matches within the 90\% radius.}
\tablenotetext{c}{Estimated number of chance background matches within the 90\% radius.}
\tablenotetext{d}{APM classification on both $E$ and $O$ plates.}
\enddata
\end{deluxetable}

\begin{deluxetable}{llccc}
\tablewidth{0pt}
\tablecaption{Confidence and Reliability for Single Band Detections\label{table-confidence-single-band}}
\tablehead{
&\multicolumn{2}{c}{Blue-only objects}
&\multicolumn{2}{c}{Red-only objects} \\
\colhead{Radius} 
&\colhead{Completeness} 
&\colhead{Reliability} 
&\colhead{Completeness} 
&\colhead{Reliability}  \\
\colhead{(arcsec)} 
&
&
&
&
}
\startdata
   1.0   &0.56       &0.93        &0.46       &0.98 \\
   2.0   &0.78       &0.82        &0.70       &0.94 \\
   3.0   &0.88       &0.71        &0.78       &0.89 \\
   4.0   &0.94       &0.60        &0.84       &0.83 \\
   5.0   &0.97       &0.50        &0.87       &0.77 \\
\enddata
\end{deluxetable}

\begin{deluxetable}{rrrrrrrrrrrrrrrrrr}
\tabcolsep=3pt
\tabletypesize{\scriptsize}
\tablenum{4}
\tablewidth{0pt}
\tablecaption{Sample Page from FIRST-APM Source Catalog\label{table-sample}}
\tablehead{
\colhead{RA} & \colhead{Dec} & \colhead{$\Delta$RA} & \colhead{$\Delta$Dec} & \colhead{Sep} &
\colhead{Ecl} & \colhead{Epsf} & \colhead{E} & \colhead{Ocl} & \colhead{Opsf} &
\colhead{O} & \colhead{O$-$E} & \colhead{Plate} & \colhead{$S_p$} & \colhead{$S_i$} &
\colhead{Maj} & \colhead{Min} & \colhead{PA}\\
\colhead{(1)}&\colhead{(2)}&\colhead{(3)}&\colhead{(4)}&\colhead{(5)}&
\colhead{(6)}&\colhead{(7)}&\colhead{(8)}&\colhead{(9)}&\colhead{(10)}&\colhead{(11)}&
\colhead{(12)}&\colhead{(13)}&\colhead{(14)}&\colhead{(15)}&\colhead{(16)}&
\colhead{(17)}&\colhead{(18)}
}
\startdata
08 00 00.099&+32 06 09.80&$-$1.86&$-$0.23& 1.87& 1&  2.49& 20.72&$-$1& $-$0.31& 21.88& 1.16& 1344&    1.10&     1.09&  3.0& $-$2.7&173.9 \\
08 00 00.234&+43 45 30.57&$-$0.56&$-$0.21& 0.59&$-$1& $-$0.07& 20.22& 0&  0.00&$-$22.17&$-$9.99& 1329&    8.52&     9.27&  2.1&  1.0&105.7 \\
08 00 00.243&+22 26 24.80& 0.56& 0.38& 0.67& 1&  5.09& 19.77& 0&  0.00&$-$21.85&$-$9.99&  226&    1.21&     1.05&  1.5& $-$3.0&179.1 \\
08 00 00.397&+44 39 14.16& 2.00& 0.17& 2.00& 0&  0.00&$-$20.76&$-$1&  0.24& 21.31&$-$9.99& 1317&   19.48&    22.0&  2.3&  1.5& 80.0 \\
08 00 00.752&+32 46 50.87& 0.14&$-$0.77& 0.78& 1&  4.60& 19.54& 2&  3.11& 22.46& 2.91&  989&    4.20&     3.90& $-$1.1& $-$1.7& 67.1 \\
08 00 01.358&+47 01 38.37&$-$0.16&$-$0.16& 0.23& 0&  0.00&$-$20.76&$-$1&  0.97& 22.23&$-$9.99& 1317&    2.68&     3.12&  2.2&  2.2& 90.3 \\
08 00 02.263&+51 36 09.10&$-$0.31& 0.30& 0.43& 1& 13.85& 12.08& 2& 47.05& 15.00& 2.92&  985&    1.49&     1.75&  4.2& $-$2.0& 36.6 \\
08 00 02.302&+31 17 20.98& 2.27& 1.38& 2.65&$-$1& $-$0.11& 20.48& 1&  2.85& 22.21& 1.73& 1344&    2.16&     2.73&  3.3&  2.1&169.9 \\
08 00 02.515&+35 49 27.12&$-$2.02& 4.07& 4.54&$-$1& $-$1.52& 18.88&$-$1& $-$0.96& 21.40& 2.52&  989&    1.09&     0.87&  2.2& $-$3.6&132.4 \\
08 00 02.702&+22 48 54.11&$-$0.79&$-$0.66& 1.03& 2& 24.52& 17.29& 1&  3.15& 21.62& 4.33&  226&    1.87&     2.91&  5.3&  2.6& 94.6 \\
08 00 02.932&+27 17 27.51&$-$0.64& 0.46& 0.79&$-$1&  0.80& 20.48&$-$1&  0.40& 20.77& 0.29& 1344&    5.37&     5.56&  1.3&  0.6&  0.4 \\
08 00 03.250&+36 29 42.11& 0.46& 0.14& 0.48& 1&  3.68& 19.67& 0&  0.00&$-$22.18&$-$9.99&  989&   46.71&    58.72&  3.7&  1.5& 21.6 \\
08 00 03.460&+50 30 05.76& 0.39&$-$0.16& 0.42&$-$1&  2.15& 20.23& 1&  2.61& 22.41& 2.17&  985&    1.07&     0.98&  0.5& $-$2.2&  0.8 \\
08 00 04.050&+23 26 16.30&$-$2.87& 1.68& 3.32& 2& 19.81&  9.92& 2& 21.84& 11.49& 1.56&  226&    6.00&     6.23&  1.1&  1.0&180.0 \\
08 00 04.378&+46 09 59.44& 0.83& 0.08& 0.84& 1&  7.16& 18.45&$-$1&  1.89& 22.31& 3.86& 1317&    2.01&     2.76&  4.2&  2.3&134.0 \\
08 00 04.981&+28 06 08.29&$-$2.43& 0.54& 2.49& 2& 67.99& 14.50& 1& 23.40& 18.21& 3.71& 1344&    1.02&     1.55&  4.1&  3.8& 90.5 \\
08 00 05.237&+50 10 23.62&$-$0.44& 0.16& 0.47&$-$1&  1.15& 18.93& 0&  0.00&$-$22.28&$-$9.99& 1317&    3.09&     3.88&  5.3& $-$2.4& 60.1 \\
08 00 05.463&+32 11 23.10& 0.43& 0.77& 0.88& 1&  7.23& 20.30& 0&  0.00&$-$21.83&$-$9.99& 1344&    2.92&     3.14&  2.5& $-$1.2& 23.0 \\
08 00 06.509&+43 09 53.78&$-$1.72& 2.62& 3.14& 2&  8.95& 20.09& 0&  0.00&$-$22.17&$-$9.99& 1329&    2.69&     2.07& $-$2.3& $-$2.8&138.3 \\
08 00 06.555&+32 01 47.98& 0.85& 0.40& 0.94& 2& 10.27& 20.44& 0&  0.00&$-$21.83&$-$9.99& 1344&    1.12&     0.72& $-$2.0& $-$3.9&144.3 \\
08 00 07.243&+25 22 19.95& 0.36&$-$0.19& 0.41&$-$1&  0.29& 19.99&$-$1&  0.09& 20.40& 0.40&  226&   20.86&    38.93&  7.7&  2.0&127.2 \\
08 00 09.613&+38 06 50.00& 0.52& 0.72& 0.89& 1&  2.07& 21.03& 0&  0.00&$-$22.18&$-$9.99&  989&    1.05&     1.39&  5.5& $-$2.0&141.9 \\
08 00 09.892&+44 36 45.86&$-$1.85& 2.50& 3.11&$-$1&  0.11& 20.60& 0&  0.00&$-$22.28&$-$9.99& 1317&    1.21&     1.36&  6.5& $-$3.7& 24.4 \\
08 00 10.262&+41 20 36.84&$-$1.42&$-$1.72& 2.23& 0&  0.00&$-$20.84& 1&  8.15& 22.12&$-$9.99& 1329&    1.36&     0.80& $-$3.3& $-$3.6& 91.9 \\
08 00 10.571&+34 02 20.90&$-$0.19&$-$0.93& 0.95&$-$1&  1.24& 19.27& 2&  6.19& 22.30& 3.03&  989&    1.60&     3.45&  8.9&  2.7&180.0 \\
08 00 10.974&+24 09 03.42&$-$3.57&$-$2.23& 4.21& 1&  7.99& 18.30& 1&  7.09& 20.28& 1.98&  226&    2.31&     2.34&  2.5& $-$2.1&  6.0 \\
08 00 11.108&+33 14 27.82& 0.03&$-$0.71& 0.71& 1&  2.39& 21.05& 0&  0.00&$-$22.18&$-$9.99&  989&   11.21&    11.67&  1.4&  0.7&  4.3 \\
08 00 11.197&+30 31 48.00&$-$0.28&$-$1.17& 1.21& 1&  3.24& 20.41& 0&  0.00&$-$21.83&$-$9.99& 1344&   26.07&    25.93&  0.4& $-$0.7&110.6 \\
08 00 12.213&+22 49 27.46&$-$0.02& 1.88& 1.88&$-$1&  0.29& 19.97& 0&  0.00&$-$21.85&$-$9.99&  226&    1.37&     2.14&  4.3&  3.9& 87.2 \\
08 00 12.515&+26 22 16.93&$-$2.49&$-$0.59& 2.56& 0&  0.00&$-$20.26& 1&  2.03& 20.53&$-$9.99&  226&   26.25&    27.72&  1.3&  1.3&180.0 \\
08 00 12.603&+45 39 23.51& 3.34&$-$3.24& 4.66& 2&  7.07& 20.19& 0&  0.00&$-$22.28&$-$9.99& 1317&    1.97&     3.15&  4.3&  4.1&  0.2 \\
\enddata
\end{deluxetable}

\begin{deluxetable}{lcccccrrr}
\tablenum{5}
\tablewidth{0pt}
\tablecaption{Previous Optical Identifications of Faint Radio Sources\label{table-ids}}
\tablehead{
\colhead{Position} & \colhead{Area} & \colhead{$S(1.4\hbox{ GHz})$} & \colhead{Sources} &&
\multicolumn{3}{c}{Suggested Identifications} & \colhead{Reference\tablenotemark{e}}\\
\cline{6-8}
\colhead{(1950)} & \colhead{(deg$^{2}$)} & \colhead{(mJy)} & \colhead{(1--30 mJy)} &&
\colhead{POSS-I} & \colhead{Fainter} & \colhead{Total}
}
\startdata
1204+1130 &   9.0  &        &   102 &&         21 &      -- &    21 &  1 \\

1306+2900 &        &        &       &&            &         &       &    \\
1718+5000 &   5.5  &  0.6   &   240 &&         45 &     100 &   145 &  2 \\
0014+1530 &        &        &       &&            &         &       &    \\
0840+4440 &        &        &       &&            &         &       &    \\

0852+1716 &0.44\tablenotemark{a} & 0.1 & 37 && 11 &       0 &    11 &  3 \\
1300+3034 &   0.44 &  0.1   &    40 &&          7 &       5 &    18 &  4 \\
          &  \tablenotemark{b}& 0.6 & 66 &&    11 &      14 &   25  &  5 \\

0854+1730 &   9.0  &  6.0   &   130 &&         31 &       0 &   31  &  6 \\
1204+1130 &        &        &       &&            &         &       &    \\

0310+8008 &   0.13 & 1.0\tablenotemark{c} & 14 && 3 &     0 &    3  &  7 \\
1635+6620 &   0.13 &        &       &&            &         &       &    \\

0015+1533 &   0.01 & 0.3\tablenotemark{c} & 13 && 1 &     3 &    4  &  8 \\
NGP       &  12.57 &        &   168 &&         31 &      35 &   66  &  9 \\
0927+4710 &  12.0  &        &   329 &&         53 &       4 &   57  &  10, 11 \\
1416+5242 &   0.5  & 0.03\tablenotemark{c} & 6 && 1 &     1 &    2  &  12 \\
HDF       &   0.02 & 0.02\tablenotemark{d} & 5 && 0 &     4 &    4  &  13 \\
Lockman Hole& 0.35 &  0.12  &    28 &&          5 &       1 &    6  &  14 \\
Phoenix Deep& 3.0  &  0.10  &   270 &&         50 &      12 &   62  &  15, 16 \\
    \cline{2-2} \cline{4-4} \cline{6-6} \cline{8-8}
Totals:  &   53.1  &        &   1448&&         270 = 18.6\% &  & 455 & \\

This work:&4150    &  1.0   &357,985&&     65,400\tablenotemark{f} = 18.3\% &   &65,400 \\
\tablenotetext{a}{Partial overlap with the sample of reference 6}
\tablenotetext{b}{Partial overlap with sample of reference 2}
\tablenotetext{c}{6~cm survey; fluxes translated to 20~cm using $\alpha \le -0.5$}
\tablenotetext{d}{3.6~cm survey; fluxes translated to 20~cm using $\alpha \le -0.5$}
\tablenotetext{e}{References: 1. Condon, Condon, \& Hazard 1982; 2. Windhorst, Kron, \& Koo
1984; Kron, Koo, and Windhorst 1985; 3. Condon \& Mitchell 1984; 4. Mitchell \& Condon 1985; 5. Windhorst
et al.\ 1985; 6. Coleman, Condon, \& Hazard 1985; 7. Partridge, Hilldrup \&
Ratner 1986; 8. Weistrop et al.\ 1987; 9. Benn et al. 1988; 10. Condon, Dickey,
\& Salpeter 1990; 11. Lu et al.\ 1996; 12. Fomalont et al.\ 1991; 13. Richards et
al. 1998; 14. deRuiter et al.\ 1997; 15. Hopkins et al.\ 1998; 16. Georgakakis
et al.\ 1998.}
\tablenotetext{f}{65,400 is the number of 90\%-confident identifications
given in Table~2; it does not include lower confidence associations or
objects identified with double radio sources, which boost the number of
suggested identifications to $\sim73,000$.}
\enddata
\end{deluxetable}

\end{document}